\documentclass[11pt,a4paper]{article}

\usepackage[margin=1cm,top=1.5cm,bottom=1.5cm]{geometry}

\usepackage{amsthm}

\usepackage{amssymb}
\usepackage{amsmath}
\usepackage{float}
\usepackage{natbib,hyperref}
\usepackage{scalefnt}
\usepackage{amsfonts,booktabs}
\usepackage{algorithmicx}
\usepackage{algorithm,algpseudocode}
\usepackage{setspace}
\usepackage{microtype}      
\usepackage{graphicx,subfig}
\usepackage{tikz}
\usetikzlibrary{arrows.meta, positioning, shapes, calc, fit}

\newtheorem{theorem}{Theorem}
\newtheorem{proposition}[theorem]{Proposition}%

\makeatletter
\newcommand{\keywords}[1]{%
  \vspace{2ex}
  \noindent\textbf{Keywords:} #1
  \vspace{2ex}
}
\makeatother

\begin{document}

\title{On robust Bayesian causal inference}

\author{
  Angelos Alexopoulos\thanks{Department of Economics, Athens University of Economics and Business, Greece. Email: \texttt{angelos@aueb.gr}}%
  \and
  Nikolaos Demiris\thanks{Department of Statistics, Athens University of Economics and Business, Greece. Email: \texttt{nikos@aueb.gr}}%
}

\date{} 

\maketitle

\vspace{1em}

\abstract{
This paper develops a Bayesian framework for robust causal inference from longitudinal observational data. Many contemporary methods rely on structural assumptions, such as factor models, to adjust for unobserved confounding, but they can lead to biased causal estimands when mis-specified. We focus on directly estimating time--unit--specific causal effects and use generalised Bayesian inference to quantify model mis-specification and adjust for it, while retaining interpretable posterior inference. We select the learning rate~$\omega$ based on a proper scoring rule that jointly evaluates point and interval accuracy of the causal estimand, thus providing a coherent, decision-theoretic foundation for tuning~$\omega$. Simulation studies and applications to real data demonstrate improved calibration, sharpness, and robustness in estimating causal effects.
}



\maketitle

\section{Introduction}

\label{section:intro}

\textbf{Problem setting} This paper is concerned with causal inference that is robust to model mis-specification. We posit a Bayesian approach to the generic statistical problem of potentially correcting for using the wrong model and then use this method for causal inference. Our main focus lies upon robust causal inference from time-series observational data on multiple units. Such data are common in many scientific disciplines where randomized experiments are often infeasible. We suggest to (i) start with a model that is considered the most appropriate for tackling this problem, (ii) allow for this model to be wrong, (iii) aim to understand the level of mis-specification by focusing on the appropriate estimand and (iv) derive suitably corrected causal effects. Although model mis-specification is common in causal inference, inference in such settings represents a general statistical problem, appearing for example when approximations are derived due to computational reasons, like in Variational Bayes or Composite Likelihood methods.

Recent methods in causal inference adjust for unobserved heterogeneity and confounding via data-generating processes such as factor models, see for example \cite{athey2021matrix} and \cite{xu2017generalized}. A proposal for positing models of this kind (up to transformations) as a general data-generating process may be found in \cite{Manifold}. While such models appear like a reasonable baseline they remain subject to mis-specification due to issues like missing confounders, incorrect number of latent factors or unnecessary linearity assumptions. Such mis-specification can lead to biased or misleading causal estimates.

\textbf{Methodological approach} In this paper we set the estimand of interest to be the causal effect(s) and aim to understand the degree of model mis-specification with respect to the accuracy of estimating this causal effect. We focus on factor analysis-type models which are commonly used to evaluate real-world interventions while still adjusting for known confounders. We adopt the generalized Bayesian inference paradigm and replace the likelihood with a loss function while retaining Bayesian updating and access to the posterior of the (generally asymmetric, e.g. \cite{ClaxtonNICE}) decision space. Estimation accuracy is assessed in terms of calibration and sharpness. Note that focused inference is not new in statistics, e.g. \cite{Claeskens2003}, but is integral in generalised/robust Bayesian analysis, see for example \cite{bissiri2016general} and \cite{manrtingalePost}. We select the learning rate~$\omega$ via a proper scoring rule that quantifies the joint accuracy of the posterior mean and credible interval of the causal estimand. The propriety of the scoring rule implies it favours posteriors whose summaries are correctly calibrated under the true data-generating process, providing a transparent and theoretically suitable criterion for tuning~$\omega$.  This approach aligns the choice of~$\omega$ directly with inferential accuracy rather than predictive or marginal likelihood objectives.

The proposed method has several advantages. It enables model-robust causal learning by adaptively adjusting influence of the two model components, the prior and the likelihood. It introduces (semi-)automatic bias and variance corrections based on the degree of model mis-specification. In particular, the posterior behavior adapts by either (i) shifting toward the prior when uncertainty is high or (ii) concentrating more closely to the data when appropriate. Although motivated by factor models, the method is broadly applicable to a range of causal models and other imperfect likelihood settings.

\textbf{Experiments} The method is studied in two simulation settings. First, in linear-type models where we find that our method naturally reduces to standard Bayesian inference when the model is well-specified and adapts appropriately as necessitated by the distinct types of model mis-specification. The second class of simulation experiments focus on Factor models where we also observe that selecting the optimal learning rate enables suitable location and scale corrections of the causal treatment effects. In summary, we demonstrate that the proposed method provides robustness and improvement over current alternatives.

We evaluate the proposed robust Bayesian causal inference (RBCI) framework using three real-world panel datasets. First, we revisit the California tobacco control program, a benchmark example in synthetic control evaluation \citep{abadie2010synthetic}. Second, we analyze the effect of a spatially targeted industrial policy on French regional employment \citep{gobillon2016regional}. Third, we study a digital tax enforcement intervention in the Greek energy market introduced by the Independent Authority for Public Revenue (IAPR). Across all applications, we compare against state-of-the-art matrix completion methods and find that accounting for model mis-specification leads to more accurate causal estimates and substantially improved uncertainty quantification.

\subsection{Connections to the relevant literature}

A vast literature exists on accounting for model mis-specification in general inference settings. Early work includes \cite{Huber1967} and also \cite{White1982} who showed that the asymptotic density of the maximum likelihood estimator in mis-specified models is Gaussian with sandwich covariance matrix and centered on the Kullback–Leibler divergence minimising a pseudo-true parameter value. \cite{Muller2013} studied the asymptotic behavior of the posterior in mis-specified parametric models and showed that generally it can have lower asymptotic frequentist risk. 

A common way to account for model mis-specification in the causal inference literature is via the propensity score, see \cite{PropensityScore}. This approach essentially generates a second option for obtaining consistency (in addition to using the correct outcome regression model), in the case where one specifies a correct propensity score model, thus leading to the `doubly robust' terminology. A generic framework for causal inference via Inverse Probability weights has been developed by Robins and colleagues and is typically termed G-estimation; a detailed account may be found in \cite{WhatIf}.

The propensity score approach has been criticised from the Bayesian viewpoint as an unnatural way to characterise a data generating mechanism \citep{li2023bayesian}. An excellent review of Bayesian approaches to causal inference may be found in \cite{StephensBA} who explain the difficulty of casting the propensity score within Bayesian inference. The paper covers some (seemingly unnatural for Bayesians) ways developed in the literature for circumventing such issues, including cutting feedback and two-stage inference. The authors then develop a decision-theoretic approach to incorporate the propensity score within a Bayesian non-parametric (Dirichlet process-based) framework and use the Bayesian bootstrap. \cite{Antonelli} developed a Bayesian approach to doubly robust estimation with favourable frequentist properties by estimating a propensity score and an outcome model. \cite{RayVaart} proposed to model the propensity score non-parametrically and showed that standard Gaussian process priors satisfy a semi-parametric Bernstein–von Mises theorem. Applications of Bayesian causal inference may be found in numerous diverse fields, including on algorithmic decisions (e.g. \cite{safePolicy} and \cite{algorithmAssisted}).

A parallel strand of fruitful, mostly non-Bayesian, recent developments may be found in the intersection of (bio)statistics, machine learning and econometrics, see for example \cite{DoubDebiasEcon}, \cite{NEURIPS2021_bf65417d} and also \cite{TargetLearn} where a detailed account of the targeted minimum loss estimation may be found. Recent studies include \cite{DornJasa}, \cite{GhoshRothe} who present robust confidence intervals for the Average Treatment Effect (ATE), \cite{Sarganis} who show optimality of the celebrated and widely used doubly robust estimators for both the ATE and the Average Treatment Effect on the Treated (ATT) and \cite{AugmBalWeights} who demonstrate that a type of  doubly robust estimators termed augmented balancing weights may achieve covariate balance directly instead of inverting the propensity score, see also \cite{WeightAdjustment}.

Our approach broadly follows a distinct line of research originating in Bayesian methods that aim for inference which is robust (semi-parametric-type), Bayesian and does not follow the usual Dirichlet process or Gaussian process-based Bayesian methods. Early such examples may be found in \cite{Shaun} and \cite{KenJasa} while \cite{bissiri2016general} develop a general approach that we present in detail in Section \ref{section:RobCausInf}. This method allows for model mis-specification and its degree is estimated at \cite{lyddon2019general} by matching the covariance matrix to its asymptotic limit while \cite{syring2019calibrating} estimate it by calibrating the coverage of the credible intervals, see also \cite{holmes2017assigning}. The predictive performance of such methods is investigated in \cite{jeremias}. Our method may be seen as an alternative to these methods since we are concerned with a non-asymptotic viewpoint and we aim to understand the degree of model mis-specification with respect to a focused task, estimating the causal estimand of interest.

\textbf{Outline of the paper} The remainder of the paper is organised as follows. In section \ref{section:RobCausInf} we describe the methodological part of this paper and section \ref{section:CausInf} the causal learning specifics. Section \ref{section:SimStu} contains the simulation experiments demonstrating the robustness and accuracy of the proposed method. Section \ref{section:DataAn} consists of three real world applications and the paper concludes with a discussion at section \ref{section:Disc}.

\section{Robust Bayesian causal inference}
\label{section:RobCausInf}

This section introduces our RBCI framework and the proposed criterion for selecting the learning rate~$\omega$.  The method combines generalized Bayesian inference, which allows for potential model mis-specification, with a proper scoring rule designed to evaluate the posterior accuracy of the causal estimand.  The score targets the two aspects most relevant to applied causal inference, the bias of the posterior mean and calibration of credible intervals, and provides a decision-theoretic foundation for choosing~$\omega$ that ensures well-calibrated and sharp posterior inference.

\subsection{Generalised Bayesian inference}

The generalised Bayesian inference framework provides a versatile extension to classical Bayesian updating in cases where the assumed model (and likelihood) is potentially mis-specified. In the standard Bayesian paradigm, posterior inference relies on a correctly specified likelihood $p(y_{1:n} \mid \theta)$ to update prior beliefs about the parameter $\theta$ after observing data $y_{1:n}$. When this likelihood is incorrect, a situation that is ubiquitous in causal inference with complex observational data, the resulting posterior distribution may be biased or misleading. Generalized Bayesian inference replaces the likelihood with a suitably chosen loss function $L(\theta, y_{1:n})$ that measures the discrepancy between model predictions and observed data \citep{bissiri2016general}. The resulting posterior distribution, often termed Gibbs posterior, is defined as
\begin{equation}
    \label{eq:gen_posterior_section}
    \pi_{\omega}(\theta \mid y_{1:n}) \propto \pi(\theta)\exp\{-\omega L(\theta, y_{1:n})\},
\end{equation}
where $\pi(\theta)$ denotes the prior distribution and $\omega > 0$ is a learning rate parameter that controls the weight of the loss relative to the prior.

This formulation retains the Bayesian machinery for uncertainty quantification and decision-making but relaxes the need for a fully specified generative model. In particular, the choice of $L$ allows the analyst to target specific aspects of the model, such as the fit of a specific estimand, while $\omega$ acts as a tuning (hyper)parameter to mitigate the impact of model mis-specification. In well-specified models, $\omega=1$ recovers the standard Bayesian posterior while under mis-specification alternative values of $\omega$ can yield better calibrated or sharper posteriors \citep{syring2019calibrating}. In the next subsection we describe how this framework may be used for causal inference.

\subsection{A Robust Bayesian approach to causal inference}

Causal inference from observational data involves estimating causal estimands, such as average treatment effects or unit-specific intervention effects, in the presence of confounding and other complexities. Existing Bayesian approaches typically proceed by specifying a data-generating model and performing posterior inference under the assumption that the model is correct, see for example \citep{saarela2016bayesian}. However, deviations from the assumed model, e.g., unmeasured confounders, or omitted dynamics can induce bias in the resulting causal estimates. 

We aim to account for such mis-specification by (i) start with a model that may reasonably be considered the state of the art, (ii) use generalised Bayesian inference to allow for this model to be incorrect, (iii) assess the degree of mis-specification by focusing on the causal estimand $\tau$ itself and (iv) derive suitably corrected causal effects. In essence, rather than attempting to fully recover the data-generating mechanism and then estimate $\tau$ as a by-product, we take a somewhat less ambitious approach and aim to directly calibrate the posterior distribution of $\tau$, using a loss function tailored to causal estimation accuracy. This shift of focus, from the correct model to robust estimand learning, is a central characteristic of the generalized Bayesian approach and we suggest it is suitable for causal inference. 

\subsubsection{Selecting the loss function}

Since our aim is to understand the degree of model mis-specification, we select a model-based type of loss function that encompasses parametric models and can incorporate covariates. A natural approach is to adopt the negative log-likelihood of the selected \emph{baseline model} as the loss function, scaled by a learning rate $\omega$, leading to a \emph{power posterior} (or tempered likelihood). Hence, our primary formulation corresponds to
\[
L(\theta, y_{1:n}) = - \log p(y_{1:n} \mid \theta),
\]
yielding the Gibbs posterior in \eqref{eq:gen_posterior_section}. This choice preserves interpretability since it reduces to standard Bayesian inference when $\omega =1$ and allows for robust adaptation; when $\omega < 1$, the posterior distribution inflates uncertainty to account for potential model mis-specification while $\omega > 1$ leads to sharpening around the data. The choice of the likelihood function is problem-specific, effectively defined by the state of the art model for the problem at hand. The term $\omega$ is essentially a hyper-parameter that may not naturally be estimated by the data and the next section is concerned with its selection.

Interestingly, alternative loss functions that target bias reduction or variance control more directly, e.g., doubly-robust or orthogonal losses, could also be used in principle. However, these generally break conjugacy and lead to substantially higher computational cost. For this reason, we retain the likelihood-based loss and rely on $\omega$ to provide a computationally efficient form of robustness.

\subsubsection{Estimating the optimal learning rate}
\label{sec:omega_selection}

Our inference procedure defines a grid of learning-rate values
$\{\omega_1, \dots, \omega_B\}$ and selects the value that optimizes a
decision-theoretic criterion based on the accuracy of estimating the causal
effect~$\tau$. For each~$\omega$, let $\Pi_\omega(\tau \mid y_{1:n})$ denote
the marginal posterior (or posterior predictive) distribution of~$\tau$ induced
by the generalized posterior in \eqref{eq:gen_posterior_section}. Denote its
expectation and central (i.e. equal-tailed) $(1-\alpha)$ quantiles by
\[
\widehat{\tau}_\omega = \mathbb{E}_{\Pi_\omega}[\tau], \qquad
(\ell_\omega, u_\omega)
=
\big(F_{\Pi_\omega}^{-1}(\alpha/2),\,F_{\Pi_\omega}^{-1}(1-\alpha/2)\big),
\]
respectively, where $F_{\Pi_\omega}$ is the cumulative distribution function (CDF) of $\Pi_\omega(\tau \mid y_{1:n})$. We assess the quality of each posterior through a score that combines point and interval accuracy:
\begin{equation}
\label{eq:score_omega}
S(F_{\Pi_\omega}, \tau)
=
(\widehat{\tau}_\omega - \tau)^2
+
S_{\mathrm{int}}(\ell_\omega,u_\omega;\tau),
\end{equation}
where $S_{\mathrm{int}}$ is the interval score of
\citet{gneiting2007strictly},
\[
S_{\mathrm{int}}(\ell,u;\tau)
=
(u-\ell)
+ \frac{2}{\alpha}(\tau-u)\,\mathbf{1}\{\tau > u\}
+ \frac{2}{\alpha}(\ell-\tau)\,\mathbf{1}\{\tau < \ell\}.
\]
The first term in \eqref{eq:score_omega} penalizes bias of the posterior mean,
while the second measures the width and calibration of the central
$(1-\alpha)$ credible interval.  One may interpret
$S(F_{\Pi_\omega}, \tau)$ as the specialization, at $G = F_{\Pi_\omega}$, of a
general scoring rule $S(G,\tau)$ defined on distributions~$G$ for~$\tau$; see
in the supplementary material for details, including a proof of the following Proposition.

\begin{proposition}
\label{prop:proper-main}
Let $G_0$ denote the true distribution of~$\tau$.  For any distribution~$G$,
the expected score satisfies
\[
\mathbb{E}_{\tau \sim G_0}\!\big[S(G,\tau)\big]
\;\ge\;
\mathbb{E}_{\tau \sim G_0}\!\big[S(G_0,\tau)\big],
\]
with equality whenever $G$ and $G_0$ share the same mean and central $(1-\alpha)$ quantiles. 
\end{proposition}

Hence, under the true data-generating process, the score is minimised in expectation when the posterior reports the correct mean and central $(1-\alpha)$ interval of~$\tau$.  The rule is not strictly proper for the full predictive distribution since distinct posteriors with the same mean and central interval may achieve the same expected score.  This suffices in our setting, since our focus lies on accurate point estimation and well-calibrated
intervals for causal effects rather than full distributional elicitation.
Therefore, working with a proper rather than a strictly proper scoring rule
does not compromise the validity of our evaluation criterion. If one wishes to
enforce strict propriety for the full predictive distribution,
$S(F_{\Pi_\omega}, \tau)$ could be replaced or augmented by a strictly proper
score such as the continuous ranked probability score (CRPS); see in the supplementary material for a discussion.

\begin{proposition}
\label{prop:omega-equals-one}
Let $R_n(\omega) = \mathbb{E}\!\left[S(F_{\Pi_\omega}, \tau)\right]$ denote the
expected score under the true data-generating process. In the correctly
specified Gaussian model $Y_i \sim \mathcal{N}(\mu,\sigma^2)$, with estimand of interest $\tau  = \mu$, $R_n(\omega)$
is uniquely minimized at $\omega = 1$. The result holds exactly for a flat
(improper) prior on $\tau$ when $\sigma^2$ is known and asymptotically for any
regular conjugate prior on $\tau$ and $\sigma^2$ when $\sigma^2$ is unknown.
\end{proposition}
Intuitively, under correct model specification, the generalized posterior
reduces to the standard Bayesian posterior, and the expected score
$R_n(\omega)=\mathbb{E}[S(F_{\Pi_\omega},\tau)]$ attains its unique minimum at
$\omega=1$. This can be verified analytically in the normal case, where the
posterior mean is the sample mean and the posterior variance scales as
$\sigma^2/(\omega n)$, implying that the expected score is minimized when the
posterior uncertainty matches the true sampling variance. A detailed proof is
provided in the supplementary material. The optimal learning rate is then chosen as
\[
\omega^\star
=
\arg\min_{\omega \in \{\omega_1,\dots,\omega_B\}} S(F_{\Pi_\omega}, \tau),
\]
balancing sharpness and calibration of the posterior distribution of~$\tau$
against the accuracy of its mean. For comparison, one may consider the mean-squared-error criterion
\begin{equation}
\label{eq:score_mse}
S_{\mathrm{MSE}}(\omega,\tau)
=
\big(\mathbb{E}_{\Pi_\omega}[\tau] - \tau\big)^2
+ \mathrm{Var}_{\Pi_\omega}(\tau),
\end{equation}
which decomposes uncertainty into squared bias and variance.  In simple
Gaussian cases, the interval width and posterior variance are linked via
\[
u_\omega - \ell_\omega
=
2 z_{1-\alpha/2}\sqrt{\mathrm{Var}_{\Pi_\omega}(\tau)}
=
2 z_{1-\alpha/2}\,\frac{\sigma}{\sqrt{\omega n}},
\]
for some~$\sigma$, but in general
$S_{\mathrm{int}}(\ell_\omega,u_\omega;\tau)$ and
$\mathrm{Var}_{\Pi_\omega}(\tau)$ capture different aspects of calibration and
sharpness.

The optimization of~$\omega$ is carried out by evaluating the score
\eqref{eq:score_omega} (or, alternatively, \eqref{eq:score_mse}) over a grid of
candidate values and selecting the minimizer.  The procedure is summarized
below. Importantly, since the score in~\eqref{eq:score_omega} is proper, this procedure provides a theoretically grounded and interpretable way of tuning~$\omega$ to achieve
robust, well-calibrated inference for causal effects.

\begin{algorithm}[H]
\caption{Selecting the optimal learning rate $\omega^\star$}
\label{alg:omega_selection}
\begin{algorithmic}[1]
\State Specify a grid of candidate values $\{\omega_1,\dots,\omega_B\}$.
\For{$b = 1,\dots,B$}
    \State Compute the generalized posterior $\Pi_{\omega_b}(\tau \mid y_{1:n})$.
    \State Extract $\widehat{\tau}_{\omega_b}$, $\ell_{\omega_b}$, and $u_{\omega_b}$.
    \State Evaluate the score $S(F_{\Pi_{\omega_b}},\tau)$ from~\eqref{eq:score_omega}.
\EndFor
\State Select $\omega^\star = \arg\min_{\omega_b} S(F_{\Pi_{\omega_b}},\tau)$.
\end{algorithmic}
\end{algorithm}

\section{Causal Inference}
\label{section:CausInf}

The goal of causal inference is to estimate the effect of a treatment or intervention on an outcome of interest while accounting for confounding and other sources of bias. The standard formalisation is the \emph{potential outcomes framework} \citep{rubin1974estimating}. For each unit $i$, let $Y_i(1)$ and $Y_i(0)$ denote the potential outcomes under treatment and control, respectively. The individual treatment effect is $\tau_i = Y_i(1) - Y_i(0)$ and because we only ever observe one of these potential outcomes, causal inference is fundamentally a missing data problem. A key estimand of interest is the ATE, $\tau_{\mathrm{ATE}} = \mathbb{E}[Y(1) - Y(0)]$, which measures the expected difference in outcomes if the entire population were treated versus the scenario where none were treated. In many observational and policy evaluation contexts, a more relevant estimand is the ATT, $ \tau_{\mathrm{ATT}} = \mathbb{E}[Y(1) - Y(0) \mid D=1]$
which measures the average causal impact of the treatment on those units that actually received it. 

\subsection{Cross-sectional data: regression-based approaches}

When data consist of independent observations taken at a single time point (cross-sectional data), causal effects are often estimated via outcome regression models or propensity score methods. A common approach is to specify a linear regression model:
\begin{equation}
    \label{eq:linear_regression}
    Y_i = \alpha + \tau D_i + X_i^\top \beta + \varepsilon_i, \quad \varepsilon_i \sim \mathcal{N}(0,\sigma^2),
\end{equation}
where $D_i$ is a binary treatment indicator, $X_i$ are observed covariates, and $\tau$ captures the average effect of treatment. Under the \emph{conditional unconfoundedness} assumption
\[
\{Y_i(1), Y_i(0)\} \perp D_i \mid X_i,
\]
the ordinary least squares estimator of $\tau$ identifies the ATE. Bayesian approaches proceed by specifying priors over $(\alpha, \beta, \tau, \sigma^2)$ and computing the posterior distribution \citep{saarela2016bayesian}. Extensions include flexible outcome regression, propensity score weighting, and doubly robust estimators. The strength of these approaches lies in their simplicity and interpretability. However, they rely critically on observing all relevant confounders and correctly specifying the outcome model. When these conditions fail — for instance, if unobserved heterogeneity influences both treatment and outcomes — the estimated causal effects will likely be biased.

\subsection{Panel data: latent factor and synthetic control models }

\begin{figure}[t]
  \centering
   \includegraphics[scale=0.2]{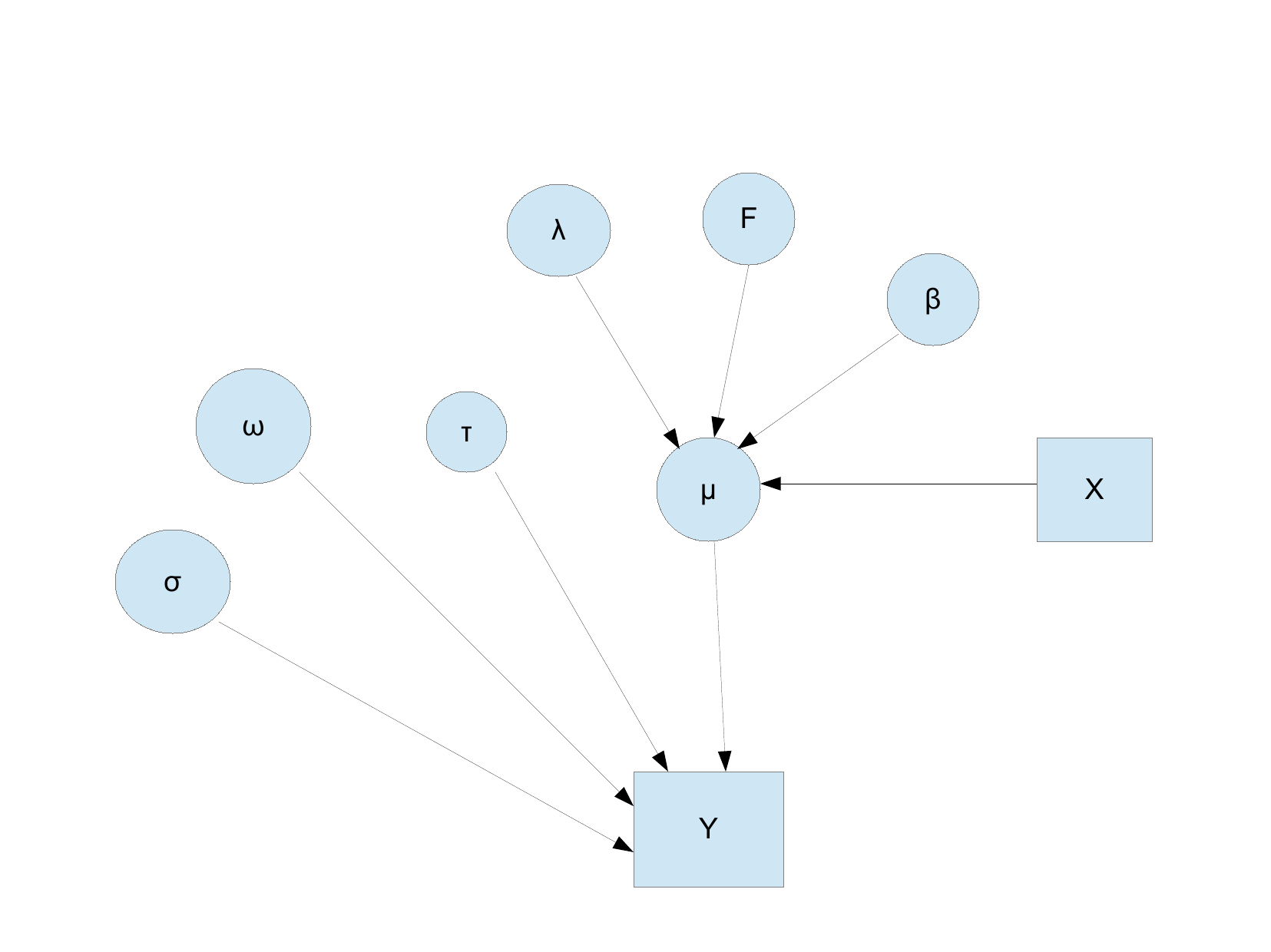}
\caption{
Directed acyclic graph representing the structure of the generalized
Bayesian causal model. The treatment effect $\tau$ and noise scale $\sigma$
govern the outcome $Y$, while latent factors $F$ with loadings $\lambda$
generate the baseline structure. Covariates $X$ influence both the latent
structure and outcome through regression coefficients $\beta$. The learning
rate $\omega$ modulates the likelihood contribution in the generalized
posterior, and $\mu$ denotes prior location hyperparameters.
}
  \label{fig:DAG}
\end{figure}

In many applied settings, interventions are introduced at different times across units, a situation known as \emph{staggered adoption}. Extending the potential outcomes framework \citep{rubin1974estimating} to longitudinal data, we define for each unit $i$ and time $t$ the potential outcomes $Y_{it}(1)$ and $Y_{it}(0)$ under treatment and control, respectively. The causal effect of interest at the unit--time level is
\[
\tau_{it} = Y_{it}(1) - Y_{it}(0),
\]
but as in the cross-sectional setting, we only observe one of the two potential outcomes at any given time, making causal inference a missing data problem. Specifically, let $T_i$ be the time treatment begins for unit $i$; for $t \ge T_i$ we observe $Y_{it}(1)$ and must infer the missing counterfactual $Y_{it}(0)$. The ATT and its dynamic form can then be written as
\begin{equation*}
\label{eq:att_staggered}
\begin{aligned}
\tau_{\mathrm{ATT}}
&= \frac{1}{|\mathcal{T}|} \sum_{t \in \mathcal{T}} \tau_{\mathrm{ATT}}(t)
= \frac{1}{|\mathcal{T}|} \sum_{t \in \mathcal{T}} 
      \left( \frac{1}{|\mathcal{I}_t|} \sum_{i \in \mathcal{I}_t}
      \big[ Y_{it}(1) - Y_{it}(0) \big] \right),
\end{aligned}
\end{equation*}
where $\mathcal{I}_t = \{ i : T_i \le t \}$ is the set of units treated by time $t$ and $\mathcal{T}$ denotes the set of post-treatment periods. These dynamic effects capture how the impact of treatment evolves with exposure length. A widely used modelling strategy for staggered adoption settings, following \citet{xu2017generalized}, specifies the observed outcome as
\begin{equation}
\label{eq:factor_model_observed}
Y_{it} = \tau_{it} D_{it} + \lambda_i^\top f_t + X_{it}^\top \beta + \varepsilon_{it},
\end{equation}
where $D_{it}$ is a binary treatment indicator, $\lambda_i$ are unit-specific factor loadings, $f_t$ represent time-varying latent factors common across units, $X_{it}$ are observed covariates with coefficient vector $\beta$, and $\varepsilon_{it}$ is an idiosyncratic error. This specification allows for both unobserved heterogeneity and common shocks affecting units through the interactive fixed effects $\lambda_i^\top f_t$. A graphical representation of the models defined in \ref{eq:factor_model_observed} (and \ref{eq:linear_regression}) is depicted in Figure \ref{fig:DAG} where we group the mean behaviour in $\mu$ and split the uncertainty/structural parameters affecting the response data Y.

Under the assumption that treatment affects outcomes only through $\tau_{it} D_{it}$, the untreated potential outcomes follow
\begin{equation}
\label{eq:factor_model_staggered}
Y_{it}(0) = \lambda_i^\top f_t + X_{it}^\top \beta + \varepsilon_{it},
\end{equation}
which corresponds to the data-generating process for control units ($D_{it} = 0$). Estimation of $\lambda_i$ and $f_t$ from pre-treatment data enables counterfactual prediction of $Y_{it}(0)$ in post-treatment periods, forming the basis of generalized synthetic control and related matrix completion approaches \citep{xu2017generalized,athey2021matrix}.

Conditional on estimates $\widehat{Y}_{it}(0)$ of untreated potential outcomes, we can write the overall and time-specific ATT in a single numbered display
\begin{equation}
\label{eq:att_staggered_hat}
\begin{aligned}
\widehat{\tau}_{\mathrm{ATT}}
&= \frac{1}{|\mathcal{T}|} \sum_{t \in \mathcal{T}} \widehat{\tau}_{\mathrm{ATT}}(t)
= \frac{1}{|\mathcal{T}|} \sum_{t \in \mathcal{T}}
      \left( \frac{1}{|\mathcal{I}_t|} \sum_{i \in \mathcal{I}_t}
      \big[ Y_{it} - \widehat{Y}_{it}(0) \big] \right).
\end{aligned}
\end{equation}
Despite their flexibility and empirical success, latent factor and matrix completion approaches rely on structural assumptions that may be violated in practice, leading to biased $\widehat{Y}_{it}(0)$ and therefore distorted $\tau_{\mathrm{ATT}}$ and $\widehat{\tau}_{\mathrm{ATT}}(t)$ estimates. 

\subsection{Baseline Likelihood and Estimation Methodology}
 
Cross-sectional and panel data are widely analysed using frequentist, ML and
Bayesian approaches. In cross-sectional settings, causal effects are often
estimated under conditional unconfoundedness via regression or propensity score
methods \citep{hogan2004instrumental,seaman2018introduction,abadie2024doubly}.
Flexible ML approaches \citep{athey2019estimating,van2011targeted} relax
functional assumptions but typically offer weaker uncertainty quantification.

Bayesian methods provide coherent uncertainty assessment and are well-suited to
complex data structures, including latent confounding. However, robustness to
\emph{model mis-specification} has received less attention in this literature,
with some notable exceptions such as BART \citep{hill2020bayesian} and Bayesian
doubly robust estimators \citep{saarela2016bayesian}. This motivates a
generalised Bayesian approach designed to remain reliable even when the outcome
model is imperfect.

\subsubsection{Bayesian estimation in cross-sectional studies}

In the Gaussian linear model~\eqref{eq:linear_regression}, the generalized
Bayesian posterior from~\eqref{eq:gen_posterior_section} is defined in terms of
the loss function
$L(\theta,Y)$, where $\theta = (\alpha,\beta,\tau,\sigma^2)$.  We take
$L(\theta,Y)$ to be the negative log-likelihood, which (up to an additive
constant) is proportional to the usual squared-error loss:
\[
L(\theta, Y)
=
\frac{1}{2\sigma^2}
\sum_{i=1}^n
\big(Y_i - \alpha - \tau D_i - X_i^\top \beta\big)^2.
\]
Substituting this choice of $L$ into \eqref{eq:gen_posterior_section} gives
\[
\pi_{\omega}(\theta \mid Y) 
\propto 
\pi(\theta)\,
\exp\!\left\{-\frac{\omega}{2\sigma^2}\sum_{i=1}^n\big(Y_i-\alpha-\tau D_i-X_i^\top\beta\big)^2\right\}.
\]
Equivalently (up to a factor independent of $\theta$ when $\sigma^2$ is fixed),
\[
\pi_{\omega}(\theta, \sigma^2 \mid Y) 
\propto 
\pi(\theta)\,
\Big[\mathcal{N}\!\big(Y \mid \alpha\mathbf{1}+\tau D+X\beta,\;\sigma^2 I_n\big)\Big]^{\omega}
\]
\[
\propto\;
\pi(\theta)\,
\mathcal{N}\!\big(Y \mid \alpha\mathbf{1}+\tau D+X\beta,\;(\sigma^2/\omega)\, I_n\big).
\]

Clearly, the learning rate $\omega$ controls the relative weight of the data and the prior in the posterior distribution and directly influences the effective noise level in the model. Hence, in this simple case it holds that raising the Gaussian likelihood to the power $\omega$ is equivalent (up to a proportionality constant) to using a Gaussian likelihood with variance $\sigma^2 / \omega$. This means that $\omega$ rescales the variance term and thus the precision with which parameters are estimated. When $\omega = 1$, we recover standard Bayesian updating under the true model. Values $\omega < 1$ downweight the likelihood relative to the prior, effectively inflating the variance to $\sigma^2 / \omega > \sigma^2$ leading to broader credible intervals, similarly to the sandwich estimators. This type of conservative inference is desirable when the model is mis-specified, including when the likelihood is overly informative relative to the true data-generating process. Conversely, $\omega > 1$ shrinks the effective variance producing sharper posteriors. Overall, $\omega$ acts as a tuning parameter that balances robustness and efficiency: smaller values favour robustness to mis-specification by tempering the likelihood’s influence, while larger values favour efficiency by emphasising the information in the data.

\subsubsection{Panel data estimation}
\label{section:panel_estimation}

Estimation of causal effects in panel data settings with staggered treatment adoption typically relies on predicting the unobserved counterfactual outcomes $Y_{it}(0)$ using latent factor models such as \eqref{eq:factor_model_staggered}. The most common frequentist approach estimates the latent factors and loadings by least squares or principal components methods \citep{bai2009panel}, after which counterfactuals $\widehat{Y}_{it}(0)$ are constructed and plugged into \eqref{eq:att_staggered} to obtain estimates of $\tau_{\mathrm{ATT}}$ and $\tau_{\mathrm{ATT}}(t)$. Several approaches build on this framework. Specifically, synthetic control methods \citep{abadie2010synthetic} construct a weighted combination of untreated units to approximate each treated unit’s counterfactual trajectory, while generalized synthetic controls \citep{xu2017generalized} combine factor structure with observed covariates. Matrix completion approaches \citep{athey2021matrix} use low-rank structure to impute missing potential outcomes directly. Machine learning techniques such as nuclear norm minimization, regularized factor models, and neural-network-based imputations have also been proposed to improve predictive accuracy. These approaches are widely used and often perform well empirically but remain subject to model mis-specification which can bias $\widehat{Y}_{it}(0)$ and lead to misleading treatment effect estimates. Also, frequentist and ML approaches can provide limited uncertainty quantification thus complicating inference on $\tau_{\mathrm{ATT}}$ and $\tau_{\mathrm{ATT}}(t)$.

Similarly to the case of cross-sectional studies we adopt the generalised Bayesian approach to estimate the latent factor model in \eqref{eq:factor_model_staggered}. Let $Y$ the vectorised form of the \(\mathbb{R}^{N\times T}\) matrix that collects outcomes with entries \(Y_{it}\) and $\theta = (f_{1:T}, \lambda_{1:n}, \beta, \sigma^2)$ collect all unknown parameters. The Gaussian likelihood implied by this model reads
\[
p(Y \mid \theta) = \prod_{i,t} \mathcal{N}\!\left( Y_{it} \,\middle|\, \lambda_i^\top f_t + X_{it}^\top \beta,\, \sigma^2 \right).
\]

In the generalized Bayesian framework of \eqref{eq:gen_posterior_section}, we replace the likelihood with the loss
\[
L(\theta, Y) = \frac{1}{2\sigma^2} \sum_{i,t} \big( Y_{it} - \lambda_i^\top f_t - X_{it}^\top \beta \big)^2,
\]
leading to the generalized posterior
\[
\pi_\omega(\theta \mid Y) \propto \pi(\theta) \Big[ \mathcal{N}\!\left( Y \,\middle|\, \Lambda F + X\beta,\, \sigma^2 I \right) \Big]^{\omega},
\]
where $\Lambda F$ denotes the matrix of factor components with $(\Lambda F)_{it} = \lambda_i^\top f_t$. This formulation highlights the role of the scaling parameter $\omega$ and has the same interpretation as in the cross-sectional setting for $\omega < 1, =1, >1$ respectively.

Our approach relates to \citet{athey2021matrix} who select the number of latent factors by minimizing a penalised loss function; a procedure that balances model fit and complexity. Our framework can be viewed as a Bayesian generalisation of the same principle: rather than fixing the factor structure ex ante and penalising the loss function once, we adaptively calibrate $\omega$ to control the effective influence of the likelihood throughout the posterior learning process. In doing so, we address structural uncertainty (such as factor dimensionality) and provide a coherent mechanism for uncertainty quantification, also when the factor model is an approximation of the true data-generating process. Given posterior draws of $\theta$, posterior predictive samples $\widehat{Y}_{it}(0)$ are generated and used to compute $\widehat{\tau}_{\mathrm{ATT}}$ and $\widehat{\tau}_{\mathrm{ATT}}(t)$ along with their posterior distributions. This approach therefore combines coherent uncertainty quantification with robustness to model mis-specification through the calibrated choice of $\omega$.





\section{Simulation study}
\label{section:SimStu}

We now illustrate the performance of the proposed methodology through simulation experiments designed to mimic typical causal inference settings. The goal is to investigate how the proposed approach behaves under different levels of model mis-specification and to assess the effect of the learning rate $\omega$ in the two scenarios of cross-sectional regression and panel-data factor models. The details on prior specification and Bayesian computation are provided in the supplementary material.

\subsection{Simulation I: Cross-sectional regression}

Our first simulation investigates the behaviour of the generalized posterior in a regression model under varying degrees of unobserved confounding. Following \cite{luo2023semiparametric}, we generate three covariates $X_1, X_2, X_3 \overset{iid}{\sim} \mathcal{N}(0,1)$ and construct a latent confounder $U = |X_1| / \sqrt{1 - 2/\pi}$. Treatment assignment is then generated according to a logistic model 
\[
D \mid U, X_2, X_3 \sim \mathrm{Bernoulli}\!\left( \mathrm{expit}(0.4U + 0.4X_2 + 0.8X_3) \right),
\]
so that $U$ and the observed covariates jointly influence the probability of receiving treatment. Finally, outcomes are drawn from the model
\[
Y \mid D, U, X_1 \sim \mathcal{N}(\tau D + X_1 + \gamma U, 1),
\]
where the parameter $\gamma$ controls the degree of model mis-specification with larger values of $\gamma$ inducing stronger unobserved confounding and therefore greater mis-specification of the working model. We simulate $N =500$ observations from the model and assess inference regarding the causal estimand of interest, the treatment effect $\tau$. We estimate $\tau$ by targeting the generalized posterior in \eqref{eq:gen_posterior_section} with the loss function set to the baseline log-likelihood of the working model
\[
y_i = \beta_D d_i + \beta x_{i1} + \varepsilon_i, \quad \varepsilon_i \sim \mathcal{N}(0,1), i=1,\ldots,N,
\]
where $\tau = \beta_D$ and the latent confounder $U$ is omitted on purpose to introduce mis-specification. We generate data under several $\gamma$ values to vary the severity of this mis-specification.

\begin{figure}[t]
  \centering
  \subfloat(a){%
    \includegraphics[width=0.42\linewidth]{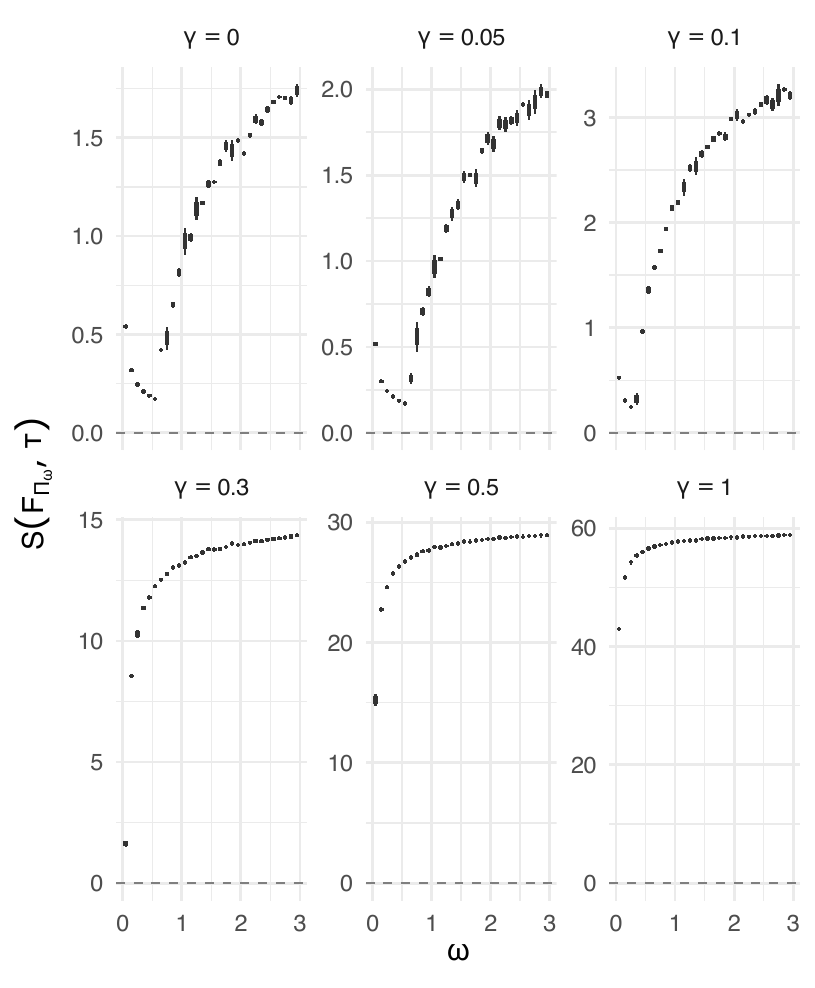}%
  }%
  \quad
  \subfloat(b){%
    \includegraphics[width=0.48\linewidth]{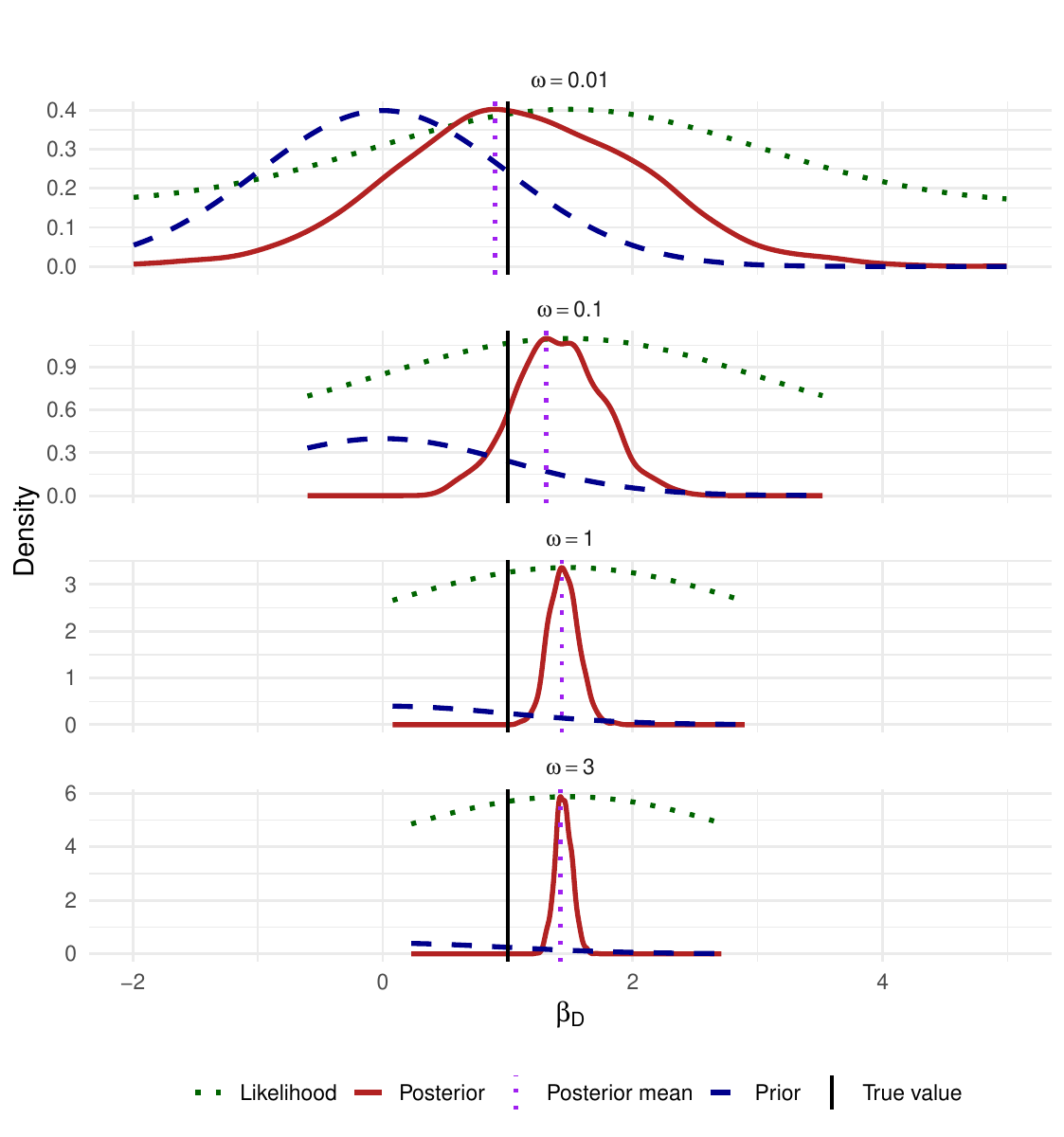}%
}
\caption{Cross-sectional regression under model mis-specification. Panel (a): score $S(F_{\Pi_\omega}, \tau)$ for different values of $\omega$ and $\gamma$; the boxplots correspond to 10 runs of the MCMC algorithm in Step 3 of Algorithm \ref{alg:omega_selection} to account for Monte Carlo variability. Panel (b): the prior/likelihood/posterior triplot of $\beta_D$ for $\gamma = 0.3$.}
  \label{fig:toy_simulation}
\end{figure}

We compute the posterior density of $\tau$ under different values of $\omega$ and evaluate the score function in \eqref{eq:score_omega}. Figure~\ref{fig:toy_simulation} illustrates two key findings. First, in the left panel it is clear that the minimiser of $S(F_{\Pi_\omega}, \tau)$ shifts systematically as the degree of model mis-specification increases. When the omitted confounding is weak (small $\gamma$), the score is minimised near $\omega = 1$, which corresponds to standard Bayesian updating and indicates that the working model is approximately correct. As the confounding strength grows, however, the optimal $\omega$ becomes smaller, downweighting the likelihood and effectively inflating posterior uncertainty. This behaviour illustrates how our framework adapts to model mis-specification: rather than over-confidently relying on a mis-specified likelihood, it automatically reduces the influence of the data and produces more conservative inference. Second, the right panel shows that this adaptive scaling has tangible benefits for inference on the treatment effect. The posterior distribution of $\beta_D$ becomes wider under mis-specification and also shifts towards the true value, correcting for the bias that would otherwise arise from ignoring the latent confounder, as appropriate. This combination of variance inflation and location adjustment leads to credible intervals with better coverage properties and more reliable causal conclusions. Together, these results highlight the key advantage of the proposed approach: by calibrating $\omega$ to the causal effect, the generalized posterior shifts towards the data-generating process, thus achieving robustness to model mis-specification without requiring explicit model corrections or complex structural assumptions.

\subsection{Simulation II: Panel data and latent factor models}
\label{sec:panel_sim}

Our second simulation examines a staggered–adoption panel-data setting with
latent confounding. We generate a $T \times N$ panel with $N=30$ units observed
over $T=100$ time periods. For each unit $i$ we draw a treatment start time
$T_i$ from a discrete uniform distribution on $\{40,\ldots,95\}$, yielding
heterogeneous post–intervention windows. The observed outcome is generated
according to
\begin{equation*}
\label{eq:sim_fac_section}
y_{it}
= \tau_{it} D_{it}
+ \lambda_i^\top f_t
+ \beta_U U_{it}(1 - D_{it})
+\epsilon_{it},
\quad
\epsilon_{it} \sim \mathcal{N}(0, 1),
\end{equation*}
where $\tau_{it}$ denotes the heterogeneous treatment effect, $D_{it} =
\mathbb{I}\{t \ge T_i\}$ is the staggered treatment indicator, and
$U_{it} = |X_{it}| / \sqrt{1 - 2/\pi}$ represents a latent source of 
time-varying outcome heterogeneity with $X_{it} \sim \mathcal{N}(0, 1)$. The parameter $\beta_U$ controls the degree of latent confounding and induces increasing model mis-specification as it grows.

\begin{figure}[t]
  \centering
  \subfloat(a){%
    \includegraphics[width=0.45\linewidth]{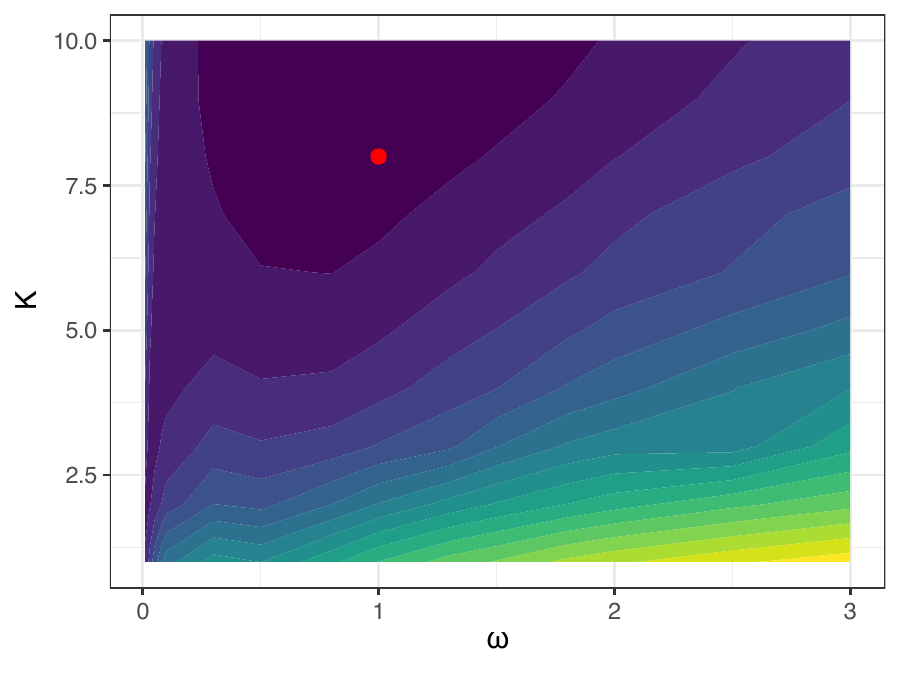}%
  }%
  \hfill
  \subfloat(b){%
    \includegraphics[width=0.45\linewidth]{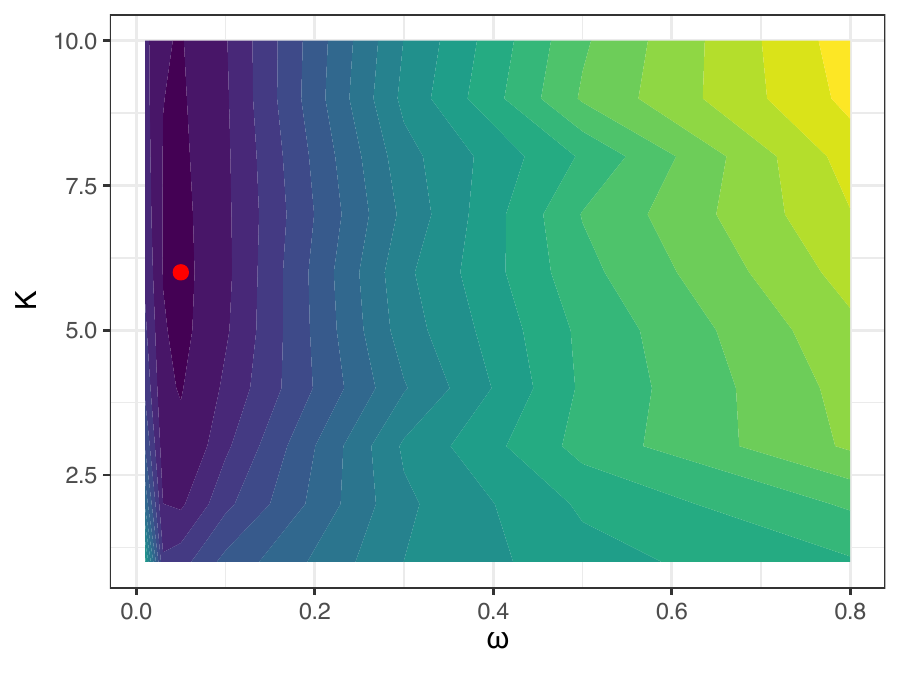}%
}
\caption{Average posterior score surface $S(F_{\Pi_\omega}, \tau)$ across a grid of values for the learning rate $\omega$ and the number of latent factors $K$. Panel (a) displays the score surface for different $\omega$ and $\gamma$ under correct model specification ($\beta_U=0$), panel (b) corresponds to strong mis-specification ($\beta_U=1$). The location of the minimum indicates the optimal balance between likelihood weighting and model complexity.}
  \label{fig:score_surface_joint}
\end{figure}

We simulate a low-rank latent structure using $K=2$ latent factors with
$\lambda_i \sim \mathcal{N}_K(0, I_K)$ and $f_t \sim \mathcal{N}_K(0, I_K)$ corresponding to the Gaussian priors specified in the supplementary material. These factors drive strong cross-sectional dependence and capture shared temporal dynamics unrelated to treatment. For treated units the counterfactual trajectories $y_{it}(0)$ are unobserved whenever $t \ge T_i$. Following the latent factor literature
\citep{athey2021matrix,samartsidis2024bayesian},
the working model for the untreated potential outcomes is specified as
\begin{equation*}
\label{eq:factor_model_sim}
y_{it}(0)
=
\lambda_i^\top f_t +\epsilon_{it}.
\end{equation*}
Figure~\ref{fig:score_surface_joint} illustrates how the average posterior score $S(\omega, K)$ varies over a grid of learning rate and factor dimension choices, under two different levels of model misspecification. The left panel corresponds to the correctly specified scenario ($\beta_U = 0$), while the right panel examines strong unobserved confounding ($\beta_U = 1$). Two notable patterns emerge: when the model is correctly specified, the score is minimised near $\omega = 1$ and at the true number of latent factors, confirming that standard Bayesian updating remains optimal in well-specified settings. By contrast, as misspecification increases, the minimiser shifts favouring $\omega < 1$ and alternate values of $K$ indicating that the generalized posterior adapts by downweighting the likelihood and adjusting model complexity. Overall, this joint surface analysis provides guidance on how the pair $(\omega, K)$ should be tuned to balance variance, bias, and complexity in different misspecification regimes.

Having established the behaviour of the score across tuning parameters, we now compare the resulting generalized posterior treatment effect estimates with those from the state-of-the-art matrix completion estimator of \citet{athey2021matrix}. We select $(K,\omega)$ using a predictive cross-validation score that targets unbiased counterfactual recovery by randomly masking $20\%$ of pre-treatment observations for never-treated units for which $\tau_{it}$ should be identically equal to zero. We vary both the degree of model mis-specification (via $\beta_U$) and the number of latent factors $K$, and, for each scenario, evaluate the minimiser of $S(F_{\Pi_{\omega}}, \tau)$ in the joint two-dimensional space. This setup allows us to assess how the proposed approach compensates for both (i) structural model errors and (ii) factor number mis-specification when estimating causal effects; see in the supplementary material for the detailed description of the adopted cross-validation procedure.
\begin{figure}[t]
    \centering
    \includegraphics[scale=0.4]{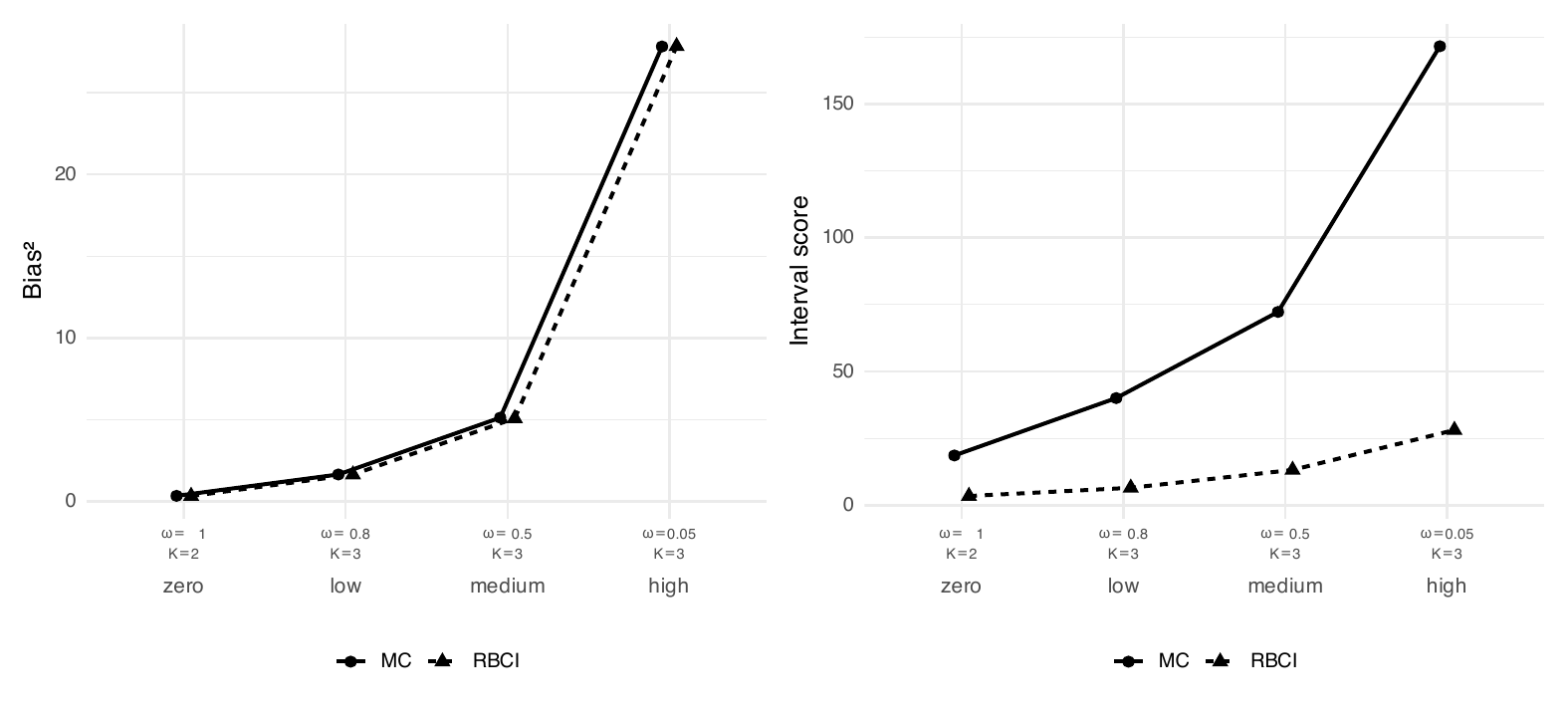}
    \caption{Comparison of squared bias (left) and interval score (right) for the proposed RBCI method and matrix completion (MC) across increasing latent factor mis-specification. Under each scenario we report the selected learning rate $\widehat{\omega}$ and number of latent factors $K^\star$.}
    \label{fig:our_vs_athey}
\end{figure}

Figure~\ref{fig:our_vs_athey} summarizes predictive causal performance under increasing levels of latent factor mis-specification. We report both the unit--time--specific squared bias of the estimated average treatment effect and the $95\%$ interval score, comparing the proposed RBCI method against matrix completion. When the factor structure is properly specified (``zero'' mis-specification), both approaches perform similarly. As mis-specification increases, however, RBCI systematically achieves lower squared bias together with substantially smaller interval scores, yielding more accurate point estimates and sharper yet well-calibrated uncertainty. The figure also displays the
selected learning rate $\widehat{\omega}$ and number of factors $K^\star$
chosen under each condition. Specifically, $\widehat{\omega}<1$ is chosen once unobserved confounding becomes more severe, tempering overconfident likelihood contributions and improving predictive robustness. These results demonstrate that adaptively tuning $(\widehat{\omega},K^\star)$ leads to more reliable
estimation of $\tau_{\mathrm{ATT}}$ without sacrificing performance when the model is correctly specified.

\section{Real data analysis}
\label{section:DataAn}

We evaluate the performance of the proposed RBCI framework on three real-world
panel datasets. In all applications we fit the latent factor model introduced
in~\eqref{eq:factor_model_observed}, with the unit--time--specific treatment
effect $\tau_{it}$ as the estimand of interest, and we estimate the generalized
posterior using the MCMC methodology described in
Section~\ref{section:panel_estimation}. The learning rate $\omega$ is selected using the same cross-validation design as in Section~\ref{section:SimStu}: we randomly mask never-treated outcomes and choose $\omega$ by evaluating how accurately the model recovers the known values of $\tau_{it}$. This is possible because $\tau_{it}$ is identically zero for all control units and for treated units prior to their intervention time, providing observed benchmarks against which the score
$S(F_{\Pi_\omega},\tau_{it})$ can be computed. 

The first application revisits the well-known California smoking intervention,
a benchmark setting for synthetic control and matrix completion methods
\citep{abadie2010synthetic,athey2021matrix}. The second examines regional
employment outcomes in France following a spatially targeted industrial policy
\citep{gobillon2016regional}. The third evaluates a digital tax enforcement
intervention introduced by the Greek IAPR in the energy market. Across all
three studies, we compare RBCI to state-of-the-art matrix completion approaches
\citep{xu2017generalized,athey2021matrix}, focusing on treatment-effect
estimation and uncertainty quantification under latent factor mis-specification.

\subsection{The California smoking intervention}
\label{subsec:Smoking}

We begin with the California tobacco control program analyzed in \citet{abadie2010synthetic}. The dataset contains annual cigarette consumption for 39 U.S. states observed over 31 years (1970--2000), with California serving as the single treated state from 1989. Following \citet{athey2021matrix}, we treat the remaining $N =38$ states as potential controls. We assess both causal point estimation and uncertainty quantification relative to matrix completion as implemented in \texttt{gsynth} \citep{gsynth2021}.

Hyperparameters in our RBCI framework are selected by minimizing a localized predictive scoring rule that combines unit--time--specific squared bias and interval score. To do so, we randomly mask $20\%$ of the outcomes belonging to never-treated states and evaluate predictions on these held-out cells, targeting unbiased out-of-sample validation. The selected configuration $(K^\star = 2,\ \widehat{\omega}=0.5)$ reflects both the extent of latent factor structure and the degree of structural mis-specification: $\widehat{\omega}<1$ indicates that a standard Bayesian update would have been overly confident and likelihood tempering yields better-calibrated uncertainty. To evaluate causal prediction performance, we follow a placebo strategy by
masking the post–1989 outcomes of a random subset (15\%) of states and treating them as pseudo-treated. Both our method and matrix completion are fit to the unmasked panel and then evaluated strictly on these masked placebo outcomes, treating them as ground truth. Full details are provided in the supplementary material.

Figure~\ref{fig:prop99-pertime} summarizes post–intervention predictive
accuracy. The left panel shows \emph{time--unit--specific} squared bias computed averaged across pseudo-treated units at each post–treatment year.  Moreover, the right panel reports the interval score for $95\%$ predictive intervals, jointly quantifying sharpness and calibration. Our approach achieves systematically lower squared bias in most post–intervention years, together with smaller interval scores, implying more
accurate and more efficient uncertainty quantification than matrix completion. These advantages reflect upon a tempered likelihood which adapts to latent factor mis-specification and produces improved time--unit–level causal estimates.

\begin{figure}[t]
\centering
\includegraphics[scale=0.3]{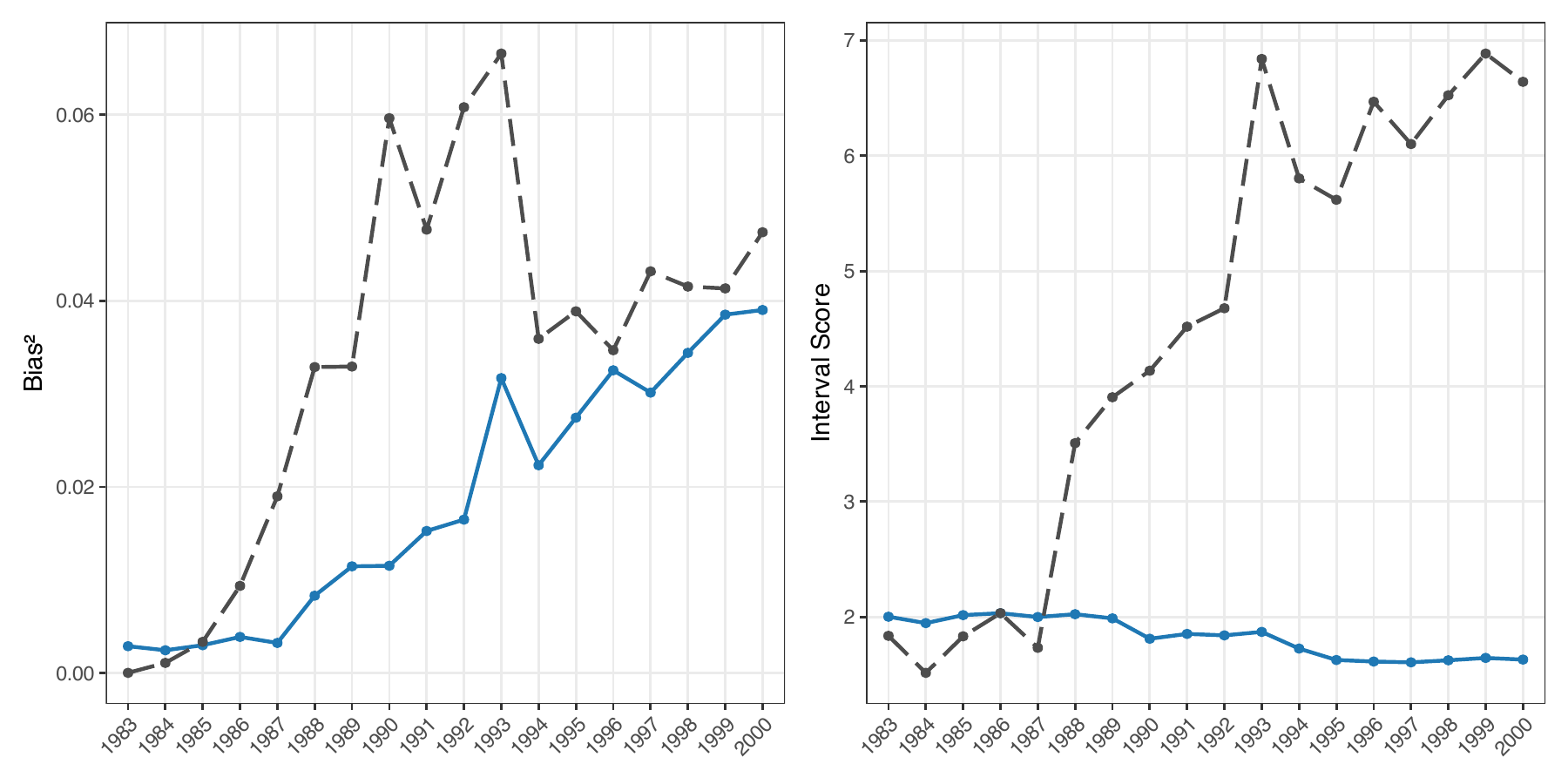}
\caption{Per-Year predictive accuracy on masked pseudo-treated  states in the California smoking study. Squared bias (left) and interval score (right), reported relative to the Proposition~99 intervention year (1989), for the proposed robust Bayesian causal inference method (solid) and matrix completion (dashed).}
\label{fig:prop99-pertime}
\end{figure}

\subsection{The French regional industrial policy}
\label{subsec:France}

We next examine the regional employment intervention studied in \citet{gobillon2016regional}, where industrial subsidies were allocated to selected French labour market areas. The dataset contains annual employment measures for $N=148$ areas observed over $T=20$ years (1997-2016), among which 13 received treatment while the remaining 135 serve as never-treated controls. 

We evaluate predictive performance using the same experimental pipeline as in Section~\ref{subsec:Smoking}. Table~\ref{tab:france_B_IS} reports results for three employment measures: firm entries, firm exits, and total employment. Our approach produces substantially improved uncertainty quantification, interval scores are reduced by $35$–$65\%$ in all cases, while achieving similar point accuracy. The benefits are most pronounced for the ``entries'' outcome, where matrix completion yields overly dispersed uncertainty intervals despite lower squared bias. Notably, for all outcomes the selected learning rate satisfies $\widehat{\omega} \ge 1$, in contrast to the California study, suggesting that the latent factor structure in this French employment panel is comparatively well specified and requires little likelihood tempering for accurate predictive uncertainty.

\begin{table}[H]
\centering
\caption{Predictive performance on masked placebo-treated regions in the French employment study. Average unit--time--specific squared bias (Bias$^2$) and $95\%$ interval score (IS) over post-intervention times for the matrix completion (MC) and the proposed robust Bayesian causal inference (RBCI) framework.}
\label{tab:france_B_IS}

\begin{tabular}{lcccccc}
\toprule
Outcome & Method & Bias$^2$ & IS & $K^\star$ & $\widehat{\omega}$ \\
\midrule
Entries & RBCI     & 0.0204 & 0.7251 & 4 & 1.5 \\
        & MC   & 0.0063 & 2.0753 & 4 & 1.5 \\
\addlinespace
Exits   & RBCI    & 0.00078 & 0.4672 & 1 & 2.0 \\
        & MC   & 0.00055 & 0.5259 & 1 & 2.0 \\
\addlinespace
Employment & RBCI   & 0.00319 & 0.4932 & 1 & 2.0 \\
           & MC & 0.00457 & 1.3319 & 1 & 2.0 \\
\bottomrule
\end{tabular}
\end{table}

\subsection{Evaluation of digital intervention in the Greek energy market}
\label{subsec:GreekEnergy}

Our third dataset is concerned with a digital tax enforcement programme by the Greek IAPR aimed at improving compliance in the energy sector. The intervention rolled out across firms and regions in phases, yielding a panel with staggered treatment adoption for $T=53$ weekly observations from $1,399$ gas stations in Greece during 2020. The outcome of interest is the log-transformed weekly fuel sales of each station. The policy is operationalised through a QR-code authentication system on fuel receipts, enabling consumers to verify in real time whether transactions have been transmitted to the tax authority’s central database, thereby increasing the perceived probability of detection for non-compliant stations and altering incentives to under-report sales. Because we observe the district of each gas station, we analyse treatment effects both at the national level and separately within 13 administrative districts. Additional data details and descriptive statistics are provided in the supplementary material.

\begin{figure}[h]
  \centering
  \includegraphics[scale=0.4]{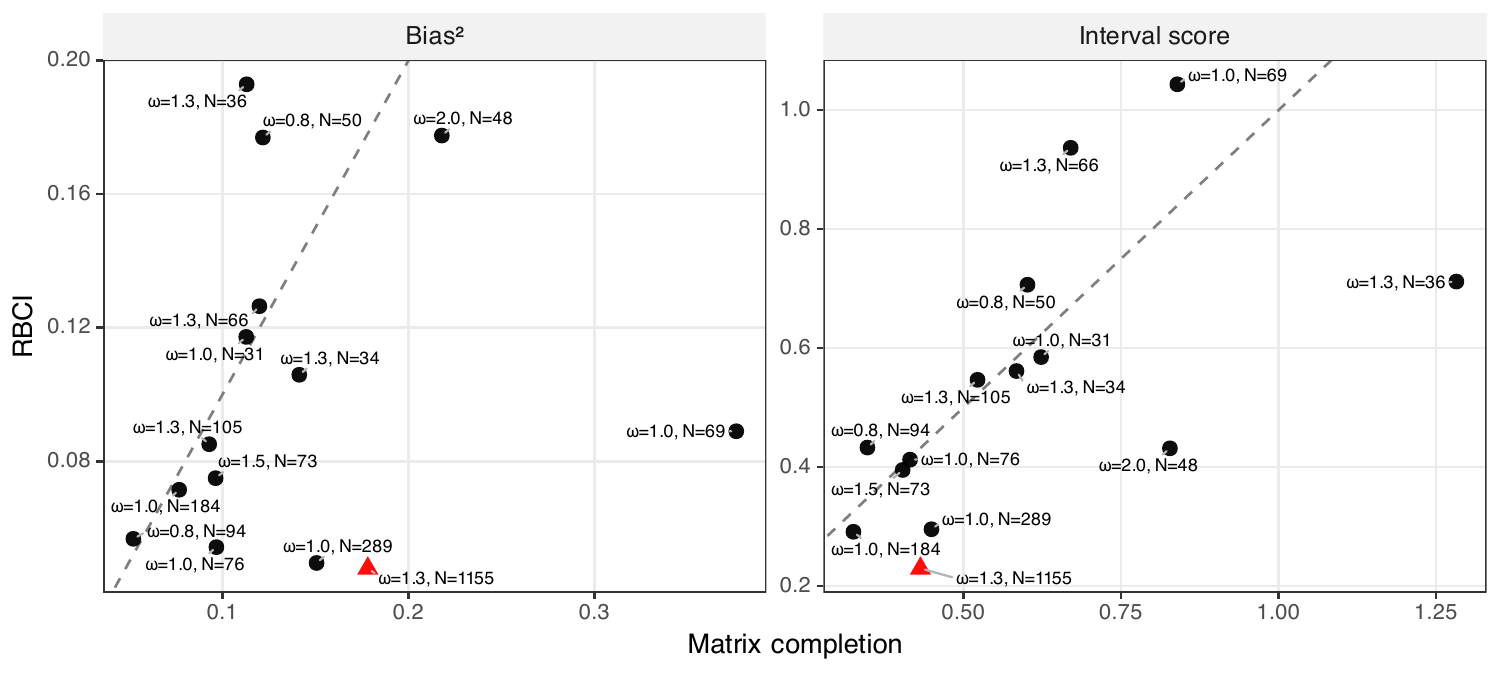}
  \caption{Each point summarises average time--unit performance for a district (Bias$^2$ on the left; interval score on the right), with labels indicating the selected learning rate $\widehat{\omega}$ and the number of observations $N$ contributing to that point. The 45$^\circ$ reference line indicates equal performance in point accuracy and uncertainty quantification between the matrix completion method (x-axis) and the proposed robust Bayesian causal inference (RBCI) framework. The analysis of the complete dataset is depicted using a red triangle.}
  \label{fig:qr-scatter}
\end{figure}

Figure~\ref{fig:qr-scatter} compares district–level and overall performance of matrix completion and RBCI. The analysis of all (N=1155) the data (depicted using a red triangle) used $\widehat{\omega}=1.3$ and resulted in improved bias and IS performance of RBCI indicating desirable point accuracy as well as sharper \emph{and} well–calibrated predictive intervals. The district-level results suggest that RBCI achieves smaller or comparable bias (with two exceptions) and typically lower interval scores. Selected learning rates vary across districts, with $\widehat{\omega}$ often in the $0.7$–$2.0$ range, reflecting heterogeneous degrees of latent–factor mis–specification: values $\widehat{\omega}<1$ temper the likelihood when the factor structure is too rigid (yielding better calibration), whereas $\widehat{\omega}\ge 1$ rates indicate that down–weighting is unnecessary and the data can be exploited more aggressively.  Overall, the Greek energy application mirrors our earlier findings: likelihood tempering adapts to structural uncertainty and delivers robust uncertainty quantification across heterogeneous panels, while remaining competitive in counterfactual point prediction.

\section{Discussion}
\label{section:Disc}

This paper postulates that causal inference based on observational data should assume the possibility for residual model error that one ought to quantify and adjust for. The robust Bayesian method outlined is broadly applicable to a range of causal models. We suggest that one should initiate this process using a model that may reasonably be considered the state of the art for the problem at hand and then allow for model mis-specification. We then adopt a non-asymptotic Bayesian approach that facilitates model correction and access to the uncertainty of the decision space. We are mostly concerned with causal inference where the issues are often structural due to missing confounders but the method is applicable to other such problems, like the presence of outliers. In addition to causal inference, the proposed framework may be used in other imperfect likelihood settings, like variational Bayes or composite likelihood cases. 

We target causal effects at the unit--time level, estimating counterfactual outcomes per treated unit and post–treatment period. We adopt a tempered posterior and select the learning rate \(\omega\) to balance calibration and sharpness. When the working model is adequate (\(\omega \approx 1\)) RBCI coincides with standard Bayesian inference while under mis-specification it adjusts the location and scale of the posterior for the unit--time–specific effects as appropriate. Our simulations and applications show that selecting \(\omega\) with a proper scoring rule—combining point accuracy and interval calibration—provides improved coverage and reduced or comparable bias compared to state-of-the-art estimators.

\textbf{Default choices and future work} A number of choices made in the present work require further discussion and exploration. The main ones are concerned with (i) the choice of loss function. This was dictated by the presence of suitable baseline models and the requirement to adjust for covariates but alternative, perhaps causality or estimand-focused, loss functions are likely suitable in different problems, including doubly-robust-like or orthogonal estimating-equation-type losses. (ii) The calibration and sharpness measures; the assessment of estimation is non-trivial in the generalised Bayes case beyond Gaussian cases and alternative accuracy measures may be operationalised by the problem, such as the nature of the estimand of interest. (iii) We found that both $\omega <1$ and $\omega >1$ may offer optimal solutions. The former offers an alternative to the tradition Bayesian prior-likelihood-posterior triplot with the posterior shifting towards the prior while the interpretation of the latter suggests `super-learning' with the posterior concentrating closer to the data and further theoretical investigation is warranted. (iv) Depending on the data, the joint posterior score space of the number of factors and $\omega$ can appear relatively flat. This space deserves further investigation and the necessary tools may be based on the Geometry of the Fisher Information and the sandwich covariance matrix. (v) We focused on tempered likelihoods using a single $\omega$ but additional flexibility may be gained using multiple, likely related, $\omega$ and insights from the weighted/composite likelihood literature may offer fruitful suggestions. (vi) Our inference procedure largely ignored the design of the intervention process. Its efficiency could further improve by adopting appropriate such designs, see for example \citet{ExpDesignCausal} and \cite{BalTrtAssign}. (vii) Our proposed framework is general and can be extended to broader classes of models, including generalised linear mixed models and survival models; this is the subject of current work. Finally (viii), recent developments (e.g. \cite{manrtingalePost}) cast Bayesian inference as a predictive task and drop the prior via a stronger focus on the estimand of interest. However, it is not immediately apparent how one may include in this framework factor-like and related general models and this remains an open problem.



\section{Acknowledgments and code}
The authors wish to thank Brian Tom and Petros Dellaportas for their valuable comments. The first author acknowledges that the research project is implemented in the framework of H.F.R.I call "Basic research Financing (Horizontal support of all Sciences) under the National Recovery and Resilience Plan “Greece 2.0" funded by the European Union – NextGenerationEU (H.F.R.I. Project Number: 15973). Replication code and data are available at
\url{https://gitlab.com/aggelisalexopoulos/robust-bayesian-causal-inference}.




\bibliographystyle{abbrvnat}
\bibliography{reference}

@article{gneiting2007strictly,
  title={Strictly proper scoring rules, prediction, and estimation},
  author={Gneiting, Tilmann and Raftery, Adrian E},
  journal={Journal of the American statistical Association},
  volume={102},
  number={477},
  pages={359--378},
  year={2007},
  publisher={Taylor \& Francis}
}

@article{gobillon2016regional,
  title={Regional policy evaluation: Interactive fixed effects and synthetic controls},
  author={Gobillon, Laurent and Magnac, Thierry},
  journal={Review of Economics and Statistics},
  volume={98},
  number={3},
  pages={535--551},
  year={2016},
  publisher={The MIT Press}
}

@article{samartsidis2024bayesian,
  title={A {B}ayesian multivariate factor analysis model for causal inference using time-series observational data on mixed outcomes},
  author={Samartsidis, Pantelis and Seaman, Shaun R and Harrison, Abbie and Alexopoulos, Angelos and Hughes, Gareth J and Rawlinson, Christopher and Anderson, Charlotte and Charlett, Andr{\'e} and Oliver, Isabel and De Angelis, Daniela},
  journal={Biostatistics},
  volume={25},
  number={3},
  pages={867--884},
  year={2024},
  publisher={Oxford University Press}
}

@article{abadie2024doubly,
  title={Doubly robust inference in causal latent factor models},
  author={Abadie, Alberto and Agarwal, Anish and Dwivedi, Raaz and Shah, Abhin},
  journal={arXiv preprint arXiv:2402.11652},
  year={2024}
}

@article{hogan2004instrumental,
  title={Instrumental variables and inverse probability weighting for causal inference from longitudinal observational studies},
  author={Hogan, Joseph W and Lancaster, Tony},
  journal={Statistical methods in medical research},
  volume={13},
  number={1},
  pages={17--48},
  year={2004},
  publisher={Sage Publications Sage CA: Thousand Oaks, CA}
}

@article{seaman2018introduction,
  title={Introduction to double robust methods for incomplete data},
  author={Seaman, Shaun R and Vansteelandt, Stijn},
  journal={Statistical science: a review journal of the Institute of Mathematical Statistics},
  volume={33},
  number={2},
  pages={184},
  year={2018}
}

@article{hill2020bayesian,
  title={Bayesian additive regression trees: A review and look forward},
  author={Hill, Jennifer and Linero, Antonio and Murray, Jared},
  journal={Annual Review of Statistics and Its Application},
  volume={7},
  number={1},
  pages={251--278},
  year={2020},
  publisher={Annual Reviews}
}

@book{van2011targeted,
  title={Targeted learning: causal inference for observational and experimental data},
  author={Van der Laan, Mark J and Rose, Sherri and others},
  volume={4},
  year={2011},
  publisher={Springer}
}

@article{athey2019estimating,
  title={Estimating treatment effects with causal forests: An application},
  author={Athey, Susan and Wager, Stefan},
  journal={Observational studies},
  volume={5},
  number={2},
  pages={37--51},
  year={2019},
  publisher={University of Pennsylvania Press}
}

@article{syring2019calibrating,
  title={Calibrating general posterior credible regions},
  author={Syring, Nicholas and Martin, Ryan},
  journal={Biometrika},
  volume={106},
  number={2},
  pages={479--486},
  year={2019},
  publisher={Oxford University Press}
}

@article{luo2023semiparametric,
  title={Semiparametric Bayesian doubly robust causal estimation},
  author={Luo, Yu and Graham, Daniel J and McCoy, Emma J},
  journal={Journal of Statistical Planning and Inference},
  volume={225},
  pages={171--187},
  year={2023},
  publisher={Elsevier}
}

@article{saarela2016bayesian,
  title={A {B}ayesian view of doubly robust causal inference},
  author={Saarela, Olli and Belzile, L{\'e}o R and Stephens, David A},
  journal={Biometrika},
  volume={103},
  number={3},
  pages={667--681},
  year={2016},
  publisher={Oxford University Press}
}

@article{abadie2010synthetic,
  title={Synthetic control methods for comparative case studies: Estimating the effect of California’s tobacco control program},
  author={Abadie, Alberto and Diamond, Alexis and Hainmueller, Jens},
  journal={Journal of the American statistical Association},
  volume={105},
  number={490},
  pages={493--505},
  year={2010},
  publisher={Taylor \& Francis}
}

@article{rubin1974estimating,
  title={Estimating causal effects of treatments in randomized and nonrandomized studies.},
  author={Rubin, Donald B},
  journal={Journal of educational Psychology},
  volume={66},
  number={5},
  pages={688},
  year={1974},
  publisher={American Psychological Association}
}

@article{xu2017generalized,
  title={Generalized synthetic control method: Causal inference with interactive fixed effects models},
  author={Xu, Yiqing},
  journal={Political Analysis},
  volume={25},
  number={1},
  pages={57--76},
  year={2017},
  publisher={Cambridge University Press}
}

@article{bai2009panel,
  title={Panel data models with interactive fixed effects},
  author={Bai, Jushan},
  journal={Econometrica},
  volume={77},
  number={4},
  pages={1229--1279},
  year={2009},
  publisher={Wiley Online Library}
}

@article{athey2021matrix,
  title={Matrix completion methods for causal panel data models},
  author={Athey, Susan and Bayati, Mohsen and Doudchenko, Nikolay and Imbens, Guido and Khosravi, Khashayar},
  journal={Journal of the American Statistical Association},
  volume={116},
  number={536},
  pages={1716--1730},
  year={2021},
  publisher={Taylor \& Francis}
}

@article{lyddon2019general,
  title={General Bayesian updating and the loss-likelihood bootstrap},
  author={Lyddon, Simon P and Holmes, CC and Walker, SG},
  journal={Biometrika},
  volume={106},
  number={2},
  pages={465--478},
  year={2019},
  publisher={Oxford University Press}
}

@article{holmes2017assigning,
  title={Assigning a value to a power likelihood in a general Bayesian model},
  author={Holmes, Chris C and Walker, Stephen G},
  journal={Biometrika},
  volume={104},
  number={2},
  pages={497--503},
  year={2017},
  publisher={Oxford University Press}
}

@article{li2023bayesian,
  title={Bayesian causal inference: a critical review},
  author={Li, Fan and Ding, Peng and Mealli, Fabrizia},
  journal={Philosophical Transactions of the Royal Society A},
  volume={381},
  number={2247},
  pages={20220153},
  year={2023},
  publisher={The Royal Society}
}

@article{bissiri2016general,
  title={A general framework for updating belief distributions},
  author={Bissiri, Pier Giovanni and Holmes, Chris C and Walker, Stephen G},
  journal={Journal of the Royal Statistical Society: Series B (Statistical Methodology)},
  volume={78},
  number={5},
  pages={1103--1130},
  year={2016},
  publisher={Wiley Online Library}
}

@manual{gsynth2021,
  title        = {gsynth: Generalized Synthetic Control Method for Causal Inference with Panel Data},
  author       = {Xu, Yiqing and Liu, Liang and Xu, M. Y.},
  year         = {2021},
  note         = {R package version 1.2.4},
  url          = {https://cran.r-project.org/web/packages/gsynth/gsynth.pdf}
}

@article{manrtingalePost,
  title={Martingale posterior distributions (with discussion)},
  author={Fong, Edwin and Holmes, Chris C and Walker, Stephen G},
  journal={Journal of the Royal Statistical Society: Series B (Statistical Methodology)},
  volume={85},
  number={5},
  pages={1357--1391},
  year={2024},
  publisher={Wiley Online Library}
}

@article{Claeskens2003,
  title   = {The Focused Information Criterion},
  author  = {Claeskens, Gerda and Hjort, Niels},
  journal = {Journal of the American Statistical Association},
  volume  = {98},
  number  = {464},
  pages   = {900--916},
  year    = {2003},
  publisher = {Taylor \& Francis}
}

@article{Manifold,
  title={Statistical Exploration of the Manifold Hypothesis (with discussion)},
  author={Whiteley, Nick and Gray, Annie and Rubin-Delanchy, Patrick},
  journal={Journal of the Royal Statistical Society: Series B (Statistical Methodology)},
  volume={to appear},
  number={},
  pages={},
  year={2025},
  publisher={Wiley Online Library}
}

@article{Huber1967,
  title={The Behavior of the Maximum Likelihood Estimates Under Nonstandard Conditions},
  author={Huber, P},
  journal={in Proceedings of the Fifth Berkeley Symposium on Mathematical Statistics and Probability.},
  volume={5th},
  number={},
  pages={221–-233},
  year={1967},
  publisher={Berkeley: University of California Press}
}

@article{White1982,
  title={Maximum Likelihood Estimation of Misspecified Models},
  author={White, H},
  journal={Econometrica},
  volume={50},
  number={1},
  pages={1--25},
  year={1982},
  publisher={Wiley Online Library}
}

@article{Muller2013,
  title={Risk of Bayesian Inference in Misspecified Models and the sandwich covariance matrix},
  author={Muller, Urlich},
  journal={Econometrica},
  volume={81},
  number={5},
  pages={1805--1849},
  year={2013},
  publisher={Wiley Online Library}
}

@article{Antonelli,
  title={Causal inference in high dimensions: a marriage between Bayesian modeling and good frequentist properties},
  author={Antonelli, Joseph and Papadogeorgou, Georgia and Dominici, Francesca},
  journal={Biometrics},
  volume={78},
  number={1},
  pages={100--114},
  year={2022},
  publisher={Wiley Online Library}
}

@article{RayVaart,
  title={Semiparametric Bayesian Causal inference},
  author={Ray, Kolyan and Van de Vaart, Aad},
  journal={The Annals of Statistics},
  volume={48},
  number={5},
  pages={2999--3020},
  year={2020},
  publisher={Institute of Mathematical Statistics}
}

@article{StephensBA,
  title={Causal Inference Under Mis-Specification: Adjustment Based on the Propensity Score (with Discussion)},
  author={Stephens, David A. and Nobre, Widemberg S. and Moodie, Erica E. M. and Schmidt, Alexandra M.},
  journal={Bayesian Analysis},
  volume={18},
  number={2},
  pages={639--694},
  year={2023},
  publisher={International Society for Bayesian Analysis}
}

@article{PropensityScore,
  title={The Central Role of the Propensity Score in Observational Studies for Causal Effects},
  author={Rosenbaum, Paul R. and Rubin, Donald B.},
  journal={Biometrika},
  volume={70},
  number={1},
  pages={41--55},
  year={1983},
  publisher={Oxford University Press}
}

@article{WhatIf,
  title={Causal Inference: What If},
  author={Hernan, Miguel A. and Robins, James M.},
  journal={},
  volume={},
  number={},
  pages={},
  year={2020},
  publisher={Boca Raton: Chapman and Hall/CRC}
}

@article{DoubDebiasEcon,
  title={Double/debiased machine learning for treatment and structural parameters},
  author={Chernozhukov, Victor and Chetverikov, Denis and Demirer, Mert and Duflo, Esther and Hansen, Christian and Newey, Whitney and Robins, James},
  journal={Econometrics Journal},
  volume={21},
  number={1},
  pages={1--68},
  year={2018},
  publisher={Oxford University Press}
}

@inproceedings{NEURIPS2021_bf65417d,
 author = {Lewis, Greg and Syrgkanis, Vasilis},
 booktitle = {Advances in Neural Information Processing Systems},
 editor = {M. Ranzato and A. Beygelzimer and Y. Dauphin and P.S. Liang and J. Wortman Vaughan},
 pages = {22695--22707},
 publisher = {Curran Associates, Inc.},
 title = {Double/Debiased Machine Learning for Dynamic Treatment Effects},
 url = {https://proceedings.neurips.cc/paper_files/paper/2021/file/bf65417dcecc7f2b0006e1f5793b7143-Paper.pdf},
 volume = {34},
 year = {2021}
}

@article{DornJasa,
  title={Doubly-valid/doubly-sharp sensitivity analysis for causal inference with unmeasured confounding.},
  author={Dorn, Jacob and Guo, Kevin and Kallus, Nathan},
  journal={Journal of the American Statistical Association},
  volume={120},
  number={549},
  pages={331--342},
  year={2025},
  publisher={Taylor and Francis}
}

@article{GhoshRothe,
  title={Assumption-robust Causal Inference},
  author={Ghosh, Aditya and Rothenhäusler, Dominik},
  journal={Arxiv},
 url = {https://arxiv.org/abs/2505.08729},
  volume={},
  number={},
  pages={},
  year={2025},
  publisher={Arxiv}
}

@article{Sarganis,
  title={Structure-agnostic Optimality of Doubly Robust Learning for Treatment Effect Estimation},
  author={Jin, Jikai and Syrgkanis, Vasilis},
  journal={COLT},
  url = {https://arxiv.org/abs/2402.14264},
  volume={},
  number={},
  pages={},
  year={2025},
  publisher={}
}

@article{AugmBalWeights,
  title={Augmented balancing weights as linear regression (with discussion)},
  author={Bruns-Smith, David and Dukes, Oliver and Feller, Avi and Ogburn, Elizabeth L},
  journal={Journal of the Royal Statistical Society: Series B (Statistical Methodology)},
  url = {https://doi.org/10.1093/jrsssb/qkaf019},
  volume={to appear},
  number={},
  pages={},
  year={2025},
  publisher={Wiley Online Library}
}

@article{Shaun,
  title={Equivalence of prospective and retrospective models in the Bayesian analysis of case‐control studies},
  author={Seaman, Shaun R. and Richardson, Sylvia},
  journal={Biometrika},
  url = {https://doi.org/10.1093/biomet/91.1.15},
  volume={91},
  number={1},
  pages={15--25},
  year={2004},
  publisher={Oxford Academic}
}

@article{KenJasa,
  title={Equivalence Between Conditional and Random-Effects Likelihoods for Pair-Matched Case-Control Studies},
  author={Rice, Kenneth},
  journal={Journal of the American Statistical Association},
  url = {https://doi.org/10.1198/016214507000001463},
  volume={103},
  number={481},
  pages={385--396},
  year={2008},
  publisher={Taylor and Francis}
}

@article{jeremias,
  title={Predictive performance of power posteriors},
  author={McLatchie, Y and Fong, E and Frazier, D T and Knoblauch, J},
  journal={Biometrika},
  url = {https://doi.org/10.1093/biomet/asaf034},
  volume={112},
  number={3},
  pages={},
  year={2025},
  publisher={Oxforf Academic}
}

@article{safePolicy,
  title={Bayesian safe policy learning with chance constrained optimization: application to military security assessment during the Vietnam War},
  author={Jia, Zeyang and Ben-Michael, Eli and Imai, Kosuke},
  journal={Journal of the Royal Statistical Society Series A: Statistics in Society},
  url = {https://doi.org/10.1093/jrsssa/qnaf122},
  volume={to appear},
  number={},
  pages={},
  year={2025},
  publisher={Oxforf Academic}
}

@article{WeightAdjustment,
  title={Direct and Stable Weight Adjustment in Non-Experimental Studies With Multivalued Treatments: Analysis of the Effect of an Earthquake on Post-Traumatic Stress},
  author={Resa, María de los Angeles and Zubizarreta, José R.},
  journal={Journal of the Royal Statistical Society Series A: Statistics in Society},
  url = {https://doi.org/10.1111/rssa.12561},
  volume={183},
  number={4},
  pages={1387-–1410},
  year={2020},
  publisher={Oxforf Academic}
}

@article{algorithmAssisted,
  title={Experimental evaluation of algorithm-assisted human decision-making: application to pretrial public safety assessment},
  author={Imai, Kosuke and Jiang, Zhichao and Greiner, D James and Halen, Ryan and Shin, Sooahn},
  journal={Journal of the Royal Statistical Society Series A: Statistics in Society},
  url = {https://doi.org/10.1093/jrsssa/qnad010},
  volume={186},
  number={2},
  pages={167-–189},
  year={2023},
  publisher={Oxforf Academic}
}

@article{ExpDesignCausal,
  title={Experimental Designs for Identifying Causal Mechanisms},
  author={Imai, Kosuke and Tingley, Dustin and Yamamoto, Teppei},
  journal={Journal of the Royal Statistical Society Series A: Statistics in Society},
  url = {https://doi.org/10.1111/j.1467-985X.2012.01032.x},
  volume={176},
  number={1},
  pages={5-–51},
  year={2013},
  publisher={Oxforf Academic}
}

@article{BalTrtAssign,
  title={Balanced and Robust Randomized Treatment Assignments: The Finite Selection Model for the Health Insurance Experiment and Beyond (with discussion)},
  author={Chattopadhyay, A. and Morris, C. N. and and Zubizarreta, J. R.},
  journal={Journal of the Royal Statistical Society Series A: Statistics in Society},
  volume={to appear},
  number={},
  pages={},
  year={2026},
  publisher={Oxforf Academic}
}

@article{ClaxtonNICE,
  title={CAUSES FOR CONCERN: IS NICE FAILING TO UPHOLD ITS RESPONSIBILITIES TO ALL NHS PATIENTS?},
  author={Claxton, Karl and Sculpher, Mark and Palmer, Stephen and Culyer, Anthony J},
  journal={Health Economics},
  url = {https://doi.org/10.1002/hec.3130},
  volume={24},
  number={1},
  pages={1-7},
  year={2015},
  publisher={Wiley Online}
}

@article{TargetLearn,
  title={Targeted Learning in Data Science: Causal Inference for Complex Longitudinal Studies.},
  author={van der Laan, Mark J. and Rose, Sherri},
  journal={},
  volume={},
  number={},
  pages={},
  year={2018},
  publisher={Springer}
}

\newpage

\begin{center}
    {\Large Appendix/Supplementary}
\end{center}

\section*{Methodology}\label{app: App1}

\subsection*{Properties of the Scoring rule $S(\tau,\omega)$}
\label{app:score-proper}

\begin{proof}[Proof of Proposition 1]
Let $G_0$ denote the true distribution of~$\tau$.  Write
$m(G)=\mathbb{E}_G[\tau]$ and let
$\ell(G)=G^{-1}(\alpha/2)$ and $u(G)=G^{-1}(1-\alpha/2)$ denote the central
quantiles.  

The squared-error term $(m(G)-\tau)^2$ is a strictly proper scoring rule for
the mean: its expected value under $G_0$ is minimized if and only if
$m(G)=m(G_0)$.  The interval score
$S_{\mathrm{int}}(\ell(G),u(G);\tau)$ is a strictly proper scoring rule for the
pair of quantiles $(\ell(G),u(G))$
\citep[][Theorem~3]{gneiting2007strictly}, so its expected value under $G_0$
is minimized if and only if
$(\ell(G),u(G))=(\ell(G_0),u(G_0))$.

For any $G$,
\[
\mathbb{E}_{\tau\sim G_0}[S(G,\tau)]
=
\mathbb{E}_{\tau\sim G_0}\big[(m(G)-\tau)^2\big]
+
\mathbb{E}_{\tau\sim G_0}\big[S_{\mathrm{int}}(\ell(G),u(G);\tau)\big].
\]
Each term is minimized at its true value, so
\[
\mathbb{E}_{\tau\sim G_0}[S(G,\tau)]
\ge
\mathbb{E}_{\tau\sim G_0}[S(G_0,\tau)],
\]
with equality if and only if $m(G)=m(G_0)$ and
$(\ell(G),u(G))=(\ell(G_0),u(G_0))$.
Thus $S(G,\tau)$ is proper.  It is not strictly proper for the full
distribution, since distinct $G$ can share the same mean and central quantiles.
\end{proof}

The proposition implies that, in expectation, the score $S(F,\tau)$ is
minimized when the reported distribution has the correct posterior mean
and central $(1-\alpha)$ credible interval of $\tau$.
Consequently, tuning $\omega$ by minimizing
$S(F_{\Pi_\omega}, \tau) = S(\Pi_\omega,\tau)$ constitutes a proper
decision-theoretic rule targeted at the inferential features of interest.

To enforce strict propriety for the full predictive
distribution, a natural alternative is the continuous ranked probability
score (CRPS),
\[
\mathrm{CRPS}(F,y)
=
\int_{-\infty}^{\infty}
\big(F(z)-\mathbf{1}\{y \le z\}\big)^2\,dz
=
\mathbb{E}_F|X-y| - \tfrac12\mathbb{E}_F|X-X'|,
\]
which is strictly proper for all univariate predictive distributions
\citep{gneiting2007strictly}.  The CRPS can be viewed as the integral of
interval scores across all nominal levels~$\alpha$ and hence enforces
strict propriety.  

\begin{proof}[Proof of Proposition 2]
Let $S(F_{\Pi_\omega},\tau) = (\widehat{\tau}_\omega-\tau)^2 + S_{\mathrm{int}}(\ell_\omega,u_\omega;\tau)$
and define the expected score
\[
R_n(\omega)\ :=\ \mathbb{E}\big[S(F_{\Pi_\omega},\tau)\big],
\]
where the expectation is with respect to the sampling distribution under the correctly specified model.

Let $\pi(\tau)\propto 1$ and Gaussian likelihood, the generalized posterior for $\tau$ is
$\mathcal{N}(\bar Y,\ \sigma^2/(\omega n))$, so $\widehat{\tau}_\omega=\bar Y$ and the equal–tailed $(1-\alpha)$ interval is
$[\ell_\omega,u_\omega]=[\bar Y\pm a_\omega]$ with $a_\omega=z_{1-\alpha/2}\,\sigma/\sqrt{\omega n}$.
Hence,
\[
\mathbb{E}\big[(\widehat{\tau}_\omega-\tau)^2\big] = \mathbb{E}[(\bar Y-\tau)^2] = \mathrm{Var}(\bar Y) = \frac{\sigma^2}{n}.
\]
Therefore,
\[
R_n(\omega)\ =\ \frac{\sigma^2}{n}\ +\ f_n(\omega),
\]
where
\[
f_n(\omega)\ :=\ \mathbb{E}\!\left[S_{\mathrm{int}}(\ell_\omega,u_\omega;\tau)\right].
\]
Let $Z=\sqrt{n}(\bar Y-\tau)/\sigma\sim\mathcal{N}(0,1)$ and
$t(\omega)=a_\omega/(\sigma/\sqrt{n})=z_{1-\alpha/2}/\sqrt{\omega}$. A standard computation yields
\[
f_n(\omega)\ =\ \frac{2\sigma}{\sqrt{n}}
\left[
t(\omega)\ +\ \frac{2}{\alpha}\Big(\phi(t(\omega)) - t(\omega)\,\bar\Phi(t(\omega))\Big)
\right],
\]
where $\phi$ and $\bar\Phi$ denote the standard normal pdf and upper tail.
Differentiating w.r.t.\ $\omega$ (using $t'(\omega)=-\,\tfrac{1}{2}t(\omega)\,\omega^{-1}$,
$\phi'(t)=-t\phi(t)$, $\bar\Phi'(t)=-\phi(t)$) gives
\[
f_n'(\omega)\ =\ -\,\frac{\sigma}{\sqrt{n}}\,
\frac{z_{1-\alpha/2}}{\omega^{3/2}}\left[\,1 - \frac{2}{\alpha}\,\bar\Phi\!\Big(\frac{z_{1-\alpha/2}}{\sqrt{\omega}}\Big)\right].
\]
Since $\bar\Phi(z_{1-\alpha/2})=\alpha/2$, we have $f_n'(1)=0$. A second–derivative check shows
$f_n''(1)>0$, hence $R_n(\omega)$ is uniquely minimized at $\omega=1$.

\paragraph{Conjugate priors and unknown variance.}
The same conclusion holds for any regular conjugate prior on~$\tau$ or when
$\sigma^2$ is unknown and assigned a normal--inverse-gamma prior. In both
settings the generalized posterior for~$\tau$ is approximately normal with mean
$m_\omega=w_\omega\bar Y+(1-w_\omega)m_0$ and variance proportional to
$\sigma^2/(\omega n)$ (conditionally on~$\sigma^2$ in the latter case).  As
$n\to\infty$, standard conjugate-updating results imply $w_\omega\to1$,
$m_\omega\overset{P}\to\bar Y$, and the posterior scale parameter converges in
probability to~$\sigma^2$. Hence the posterior mean and variance coincide
asymptotically with those under the flat-prior, known-variance case, and
$R_n(\omega)$ converges to the same limit $\sigma^2/n+f(\omega)$ whose unique
minimizer is $\omega=1$.
\end{proof}

\subsection*{Choosing $K$ and $\omega$ in panel data}
\label{app:validation}

\begin{figure}[H]
\centering
\resizebox{0.80\textwidth}{!}{%
\begin{tikzpicture}[
  node distance=1.4cm,
  box/.style={rectangle, rounded corners, draw=black, fill=gray!10,
              align=center, minimum width=5.5cm, minimum height=1cm},
  arrow/.style={->, >=latex, thick}
]

\node[box] (panel) {Observed balanced control panel};

\node[box, below left=1.6cm and 0.3cm of panel] (cv_ctrl) {
  85\% controls-only \\
  (no pseudo-treatment)
};

\node[box, below right=1.6cm and 0.3cm of panel] (test_ctrl) {
  15\% pseudo-treated controls \\
  masked post-period outcomes
};

\node[box, below=1.4cm of cv_ctrl] (sel) {
  Selection of $(K^\star, \omega^\star)$ \\
  by minimizing 
  $S(F_{\Pi_\omega}, \tau)$
};

\node[box, below=1.4cm of test_ctrl] (mask) {
  Mask only post-pseudo-treatment \\
  cells (test set)
};

\node[box, below=2.0cm of $(sel)!0.5!(mask)$] (fit) {
  Fit Bayesian factor model \\
  with $(K^\star, \omega^\star)$ \\
  (counterfactual training only)
};

\node[box, below=1.6cm of fit] (eval) {
  Time--unit--specific causal validation: \\
  $\text{Bias}^2_{it} = (\hat{\mu}_{it} - Y_{it})^2$ and IS$_{it}$ \\
  on masked pseudo-treated outcomes
};

\draw[arrow] (panel) -- (cv_ctrl);
\draw[arrow] (panel) -- (test_ctrl);
\draw[arrow] (cv_ctrl) -- (sel);
\draw[arrow] (test_ctrl) -- (mask);
\draw[arrow] (sel) -- (fit);
\draw[arrow] (mask) -- (fit);
\draw[arrow] (fit) -- (eval);

\end{tikzpicture}
}
\caption{Time--unit--specific placebo-based validation and causal evaluation. We designate 15\% of controls as pseudo-treated and mask their post-period outcomes.
Hyperparameters $(K,\omega)$ are selected to minimize local prediction error on masked cells.
The final model is evaluated by time--unit--specific squared bias and interval score.
}
\label{fig:time_unit_validation}
\end{figure}
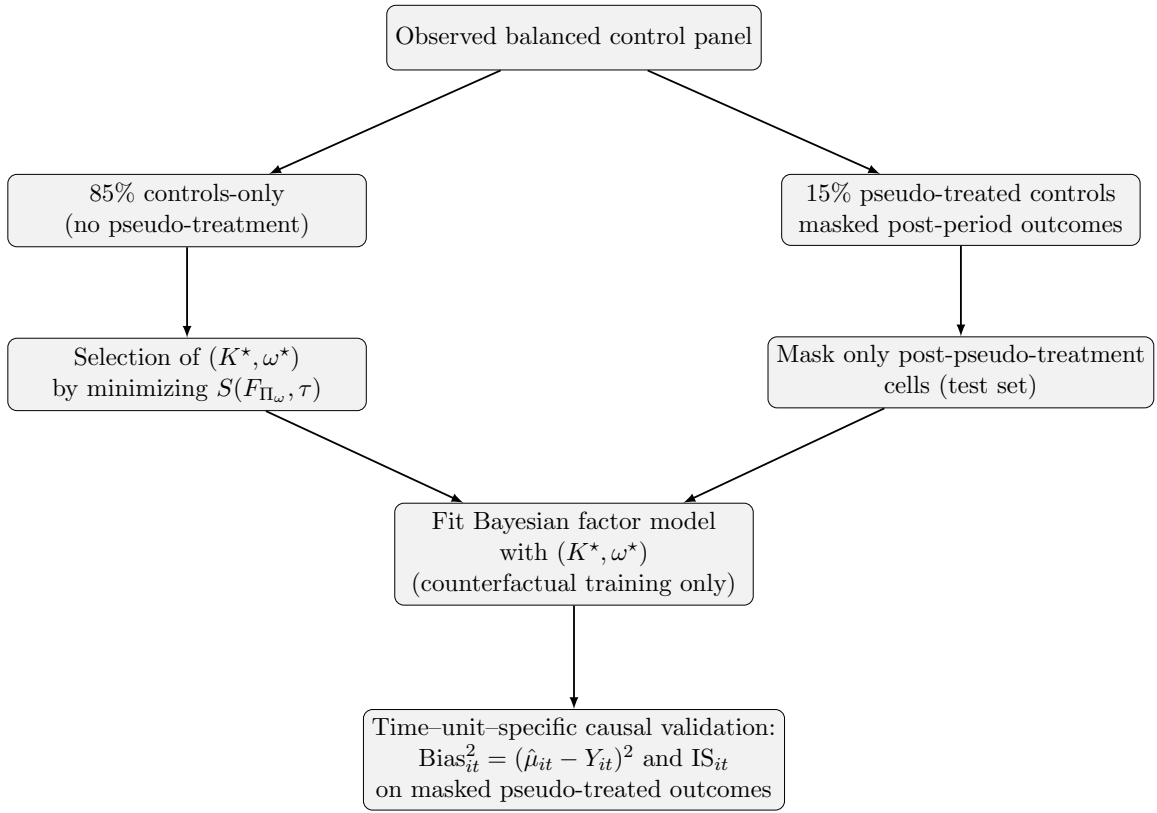

To tune the hyperparameters $(K,\omega)$ of the Bayesian factor model and assess robustness to model mis-specification, we employ a placebo-based validation design that operates at the
time--unit level as illustrated by Figure \ref{fig:time_unit_validation}. We randomly draw $15\%$ of units from this pool and assign them artificial treatment at randomly sampled post intervention times. All post-placebo outcomes for these units are removed from the data matrix and treated as missing during model fitting.
For each hyperparameter combination, the masked pseudo-treated outcomes are reconstructed solely from pre-treatment information via posterior prediction  by minimising the score  $\left(\widehat{\mu}_{it} - Y_{it}\right)^2+ \text{IS}_{it}$. Here, $\widehat{\mu}_{it}$ denotes the posterior predictive mean and thus $\left(\widehat{\mu}_{it} - Y_{it}\right)$ is its unit- time- specific bias, while $\text{IS}_{it}$ is the interval score based on full predictive intervals. The hyperparameters $(K^\star,\omega^\star)$ are selected by minimizing the average of the score across (which is an estimation of $S(F_{\Pi_\omega}, \tau)$) all masked placebo cells.

After selection, a final model is refit using the full observed pre-placebo data, and predictive performance is calculated exclusively on the held-out placebo outcomes. This procedure avoids any information leakage from masked outcomes into factor learning, jointly accounts for pointwise causal bias and uncertainty calibration, and preserves the heterogeneous, time--unit--specific structure of causal effects. Comparisons against matrix completion methods (e.g., matrix completion and generalized synthetic controls \citep{xu2017generalized, athey2021matrix}) are conducted using the exact same pseudo-treated time--unit pairs to guarantee a fair benchmark.

\section*{MCMC and modelling details}

\subsection*{Cross-sectional regression}

Following the main text, inference targets the generalized posterior for the treatment effect
$\tau$ under the \emph{working} Gaussian regression that omits the latent confounder:
\[
Y_i = \beta_D D_i + \beta X_{i1} + \varepsilon_i,\qquad
\varepsilon_i \sim \mathcal{N}(0,\sigma^2),\quad i=1,\ldots,N,
\]
with $\tau \equiv \beta_D$. (In the experiments we let $\sigma^2$ be unknown for generality; if desired it
may be fixed to $1$ to match the data generating variance.)

\paragraph{Priors.}
We place conjugate priors
\[
\begin{aligned}
(\beta_D,\beta)^\top &\sim \mathcal{N}(m_0, V_0),\\
\sigma^{-2} &\sim \mathrm{Gamma}(a_0,b_0),
\end{aligned}
\]
with hyperparameters $(m_0,V_0,a_0,b_0)$ chosen to be weakly informative.

\paragraph{Generalized (tempered) posterior.}
For a learning rate $\omega>0$, the generalized posterior is
\[
\pi_\omega(\theta\mid \mathcal{D}) \ \propto\
\Bigg\{\prod_{i=1}^N \mathcal{N}\!\big(Y_i;\, \beta_D D_i + \beta X_{i1},\ \sigma^2\big)\Bigg\}^{\omega}
\times \pi(\theta),
\qquad \theta=(\beta_D,\beta,\sigma^2).
\]
Raising the likelihood to the power $\omega$ is equivalent to inflating the effective noise from
$\sigma^2$ to $\sigma^2/\omega$ in the Gaussian kernel, thereby down-weighting the data when $\omega<1$.

\paragraph{Gibbs sampler.}
Let $X = [D\;\; X_1]$ be the $N\times 2$ design and $\beta=(\beta_D,\beta)^\top$. Conditional posteriors are standard:
\begin{enumerate}
\item $\beta \mid \sigma^2, Y \sim \mathcal{N}(m_\omega, V_\omega)$ with
\[
V_\omega \;=\; \Big(V_0^{-1} + \omega\,\sigma^{-2} X^\top X\Big)^{-1},\qquad
m_\omega \;=\; V_\omega\Big(V_0^{-1} m_0 + \omega\,\sigma^{-2} X^\top Y\Big).
\]
\item $\sigma^{-2}\mid \beta, Y \sim \mathrm{Gamma}(a_\omega,b_\omega)$ with
\[
a_\omega \;=\; a_0 + \tfrac{\omega N}{2},\qquad
b_\omega \;=\; b_0 + \tfrac{\omega}{2}\,\|Y - X\beta\|_2^2.
\]
\end{enumerate}
The sampler iterates these two updates for $M$ iterations, discarding burn-in and keeping $M_p$ posterior draws $\{\beta^{(m)},\sigma^{2(m)}\}_{m=1}^{M_p}$.
The causal parameter of interest is $\tau \equiv \beta_D$, for which we retain $\{\tau^{(m)}\}_{m=1}^{M_p}$.

From the retained draws we report:
\[
\widehat{\tau}_\omega = \mathsf{E}_m[\tau^{(m)}],\qquad
\big(\tau^{\mathrm{lo}}_\omega,\ \tau^{\mathrm{hi}}_\omega\big) = 
\big(\mathrm{q}_{0.025}(\tau^{(m)}),\ \mathrm{q}_{0.975}(\tau^{(m)})\big).
\]

For each candidate $\omega$ and each simulated dataset (indexed by $\gamma$), we compute the score
\[
\mathrm{Bias}^2_\tau(\omega) \;=\; \big(\widehat{\tau}_\omega - \tau_{\text{true}}\big)^2,\qquad
\mathrm{IS}_\tau(\omega) \;=\; 
\big(\tau^{\mathrm{hi}}_\omega - \tau^{\mathrm{lo}}_\omega\big)
+ \tfrac{2}{\alpha}\big(\tau^{\mathrm{lo}}_\omega - \tau_{\text{true}}\big)\,\mathbf{1}\{\tau_{\text{true}}<\tau^{\mathrm{lo}}_\omega\}
+ \tfrac{2}{\alpha}\big(\tau_{\text{true}} - \tau^{\mathrm{hi}}_\omega\big)\,\mathbf{1}\{\tau_{\text{true}}>\tau^{\mathrm{hi}}_\omega\},
\]
with $\alpha=0.05$. Our overall scalar score is
\[
S(\omega;\gamma)\;=\;\mathrm{Bias}^2_\tau(\omega)\;+\;\mathrm{IS}_\tau(\omega).
\]
We select $\omega^\star(\gamma)$ that minimizes $S(\omega;\gamma)$ on a grid of values. To account for Monte Carlo variability in Step 3 of Algorithm 1, we repeat the sampler $R$ times for each $(\gamma,\omega)$ and report boxplots of $S(\omega;\gamma)$ across replicates (the main text uses $R=10$).

When the working model is approximately correct ($\gamma\approx 0$), the minimizer $\omega^\star$ is typically near $1$, recovering standard Bayesian updating. As mis-specification increases ($\gamma>0$), the optimal $\omega^\star$ drifts below $1$, down-weighting the likelihood and widening the posterior for $\tau$. Empirically this reduces bias and improves the interval score, yielding more reliable uncertainty quantification for the causal effect.

\subsection*{Factor model for panel data}

For evaluation we split never-treated donor units into a \emph{train} subset (for hyperparameter calibration) and a \emph{test} subset (held out for pseudo treatment). For each test unit \(i\), we assign a staggered placebo start \(T_i\) and mask its
post-placebo outcomes
\[
\mathcal{M}\;=\;\{(i,t):\; t\ge T_i,\ i\in\text{test}\},\qquad
Y_{it}=\text{NA},\ (i,t)\in\mathcal{M}.
\]
We also define a binary placebo indicator \(Z_{it}=\mathbb{I}\{t\ge T_i,\ i\in\text{test}\}\).
When fitting, unit \(i\) contributes only its pre-placebo observations \(\{t<T_i\}\) if it is in test. Following the main text, the untreated potential outcome admits the interactive fixed-effects
representation
\[
Y_{it}(0)\;=\;\lambda_i^\top f_t\;+\;X_{it}^\top\beta\;+\;\varepsilon_{it},\qquad
\varepsilon_{it}\sim\mathcal{N}(0,\sigma_i^2),
\]
where \(f_t\in\mathbb{R}^K\) are common across units factors and \(\lambda_i\in\mathbb{R}^K\) are loadings.
We place conjugate priors
\[
f_t \sim\mathcal{N}(0,I_K),\quad
\lambda_i\sim\mathcal{N}(0,\lambda I_K),\quad
\sigma_i^{-2}\sim\mathrm{Gamma}(a,b),\quad
\beta\sim\mathcal{N}(0,\Sigma_\beta),
\]
and target the generalized (tempered) posterior with learning rate \(\omega>0\)
\[
\pi_\omega(\theta\mid Y_{\mathrm{obs}})\ \propto\
\Bigg\{\prod_{(i,t)\in\mathcal{O}}
\mathcal{N}\!\big(Y_{it}\,;\,\lambda_i^\top f_t+X_{it}^\top\beta,\ \sigma_i^2\big)\Bigg\}^{\omega}
\;\times\;\pi(\theta),
\]
where \(\mathcal{O}\) are observed (unmasked) cells and \(\theta=(\{f_i\},\{\lambda_t\},\beta,\{\sigma_i^2\})\).
Raising the likelihood to the power \(\omega\) is equivalent to inflating the effective noise
variance from \(\sigma_i^2\) to \(\sigma_i^2/\omega\) in the Gaussian terms, thus down-weighting
the likelihood when \(\omega<1\).

At iteration \(m=1,\dots,M\)
\begin{enumerate}
\item Update \(\{\lambda_i\}\) conditionally on \(\{f_t\},\beta,\{\sigma_i^2\}\) using only \((i,t)\in\mathcal{O}\).
\item Update \(\{f_t\}\) conditionally on \(\{\lambda_i\},\beta,\{\sigma_i^2\}\) using only times \(t< T_i\) if \(i\in\) test.
\item Update \(\beta\) from its Gaussian conditional.
\item Update each \(\sigma_i^2\) from its inverse-Gamma/Gamma conditional (with residual sums scaled by \(\omega\)).
\end{enumerate}
After discarding burn-in, we retain \(M_{\!p}\) draws for prediction. For each retained draw \(m\),
\[
\mu^{(m)}_{it}=\lambda_t^{(m)\top} f_i^{(m)}+X_{it}^\top\beta^{(m)},\qquad
\tilde Y^{(m)}_{it}=\mu^{(m)}_{it}+\tilde\varepsilon^{(m)}_{it},\ \ 
\tilde\varepsilon^{(m)}_{it}\sim\mathcal{N}\!\Big(0,\frac{\sigma_i^{2(m)}}{\omega}\Big).
\]
We report 
\[
\widehat{\mu}_{it}=\mathsf{E}_m[\mu^{(m)}_{it}],\quad
\widehat{Y}^{\mathrm{lo}}_{it}=\text{q}_{0.025}(\tilde Y^{(m)}_{it}),\quad
\widehat{Y}^{\mathrm{hi}}_{it}=\text{q}_{0.975}(\tilde Y^{(m)}_{it}).
\]

On the panel used to select $K$ and $\omega$ we randomly mask \(20\%\) of cells and, for each \((K,\omega)\), fit the model
to the partially observed matrix. Validation uses only masked cells \((i,t)\):
\[
\mathrm{Bias}^2=\mathsf{mean}\,(\widehat{\mu}_{it}-Y_{it})^2,\qquad
\mathrm{IS}=\mathsf{mean}\Big[(\widehat{Y}^{\mathrm{hi}}_{it}-\widehat{Y}^{\mathrm{lo}}_{it})
+\tfrac{2}{\alpha}(\widehat{Y}^{\mathrm{lo}}_{it}-Y_{it})\mathbf{1}\{Y_{it}<\widehat{Y}^{\mathrm{lo}}_{it}\}
+\tfrac{2}{\alpha}(Y_{it}-\widehat{Y}^{\mathrm{hi}}_{it})\mathbf{1}\{Y_{it}>\widehat{Y}^{\mathrm{hi}}_{it}\}\Big],
\]
with \(\alpha=0.05\). We pick \((K^\star,\omega^\star)\) minimizing \(\mathrm{Bias}^2+\mathrm{IS}\).

\subsection*{Wild bootstrap intervals for matrix completion}

The \texttt{gsynth} r-package provides counterfactual paths \(\widehat{Y}^{\mathrm{mc}}_{it}\) for treated columns by using the matrix completion framework \citep{athey2021matrix,xu2017generalized}. We construct predictive intervals on the masked test cells via a wild bootstrap procedure that preserves the panel structure and the placebo masking. Let \(\mathcal{T}\) denote the set of test units retained by \texttt{gsynth} after name-safe matching.
Define residuals on the masked test cells
\[
E_{it}\;=\;Y_{it}-\widehat{Y}^{\mathrm{mc}}_{it},\qquad (i,t)\in\mathcal{M}\cap\{i\in\mathcal{T}\}.
\]
For \(b=1,\dots,B\) (e.g.\ \(B=200\)):
\begin{enumerate}
\item Draw independent Rademacher multipliers \(\xi_i^{(b)}\in\{-1,+1\}\) for each \(i\in\mathcal{T}\),
and set \(S_{it}^{(b)}=\xi_i^{(b)}\) for all \(t\) (unit-wise sign flips).
\item Form starred treated trajectories on masked test cells:
\[
Y^{\star(b)}_{it} \;=\; \widehat{Y}^{\mathrm{mc}}_{it} + S_{it}^{(b)}\,E_{it},\quad (i,t)\in\mathcal{M}\cap\{i\in\mathcal{T}\},
\]
and replace the corresponding columns in the full controls panel \(Y\) by \(Y^{\star(b)}_{\cdot i}\)
at the \emph{same unit names}.
\item Refit \texttt{gsynth} on the starred panel (same placebo design \(Z_{it}\), same options),
retrieve \(\widehat{Y}^{\mathrm{mc},(b)}_{it}\) for \(i\in\mathcal{T}\), and extract predictions
\emph{only} at the original masked cells \((i,t)\in\mathcal{M}\cap\{i\in\mathcal{T}\}\).
\end{enumerate}
For each masked cell, take empirical quantiles over \(b=1,\dots,B\):
\[
\widehat{Y}^{\mathrm{lo}}_{it}=\mathrm{q}_{0.025}\big(\widehat{Y}^{\mathrm{mc},(b)}_{it}\big),\qquad
\widehat{Y}^{\mathrm{hi}}_{it}=\mathrm{q}_{0.975}\big(\widehat{Y}^{\mathrm{mc},(b)}_{it}\big).
\]

On the masked test set we compute
\[
\mathrm{Bias}^{2}=\mathsf{mean}\,\big(\widehat{Y}^{\mathrm{mc}}_{it}-Y_{it}\big)^2,\qquad
\mathrm{IS}^{\mathrm{mc}}=\text{IS}\big(Y_{it};\,\widehat{Y}^{\mathrm{lo}}_{it},\widehat{Y}^{\mathrm{hi}}_{it}\big),
\]
and their per-time averages across units with \(t\ge T_i\). Name-safe alignment (by unit labels)
is enforced in every bootstrap refit and extraction step to prevent column-order drift.

\section*{Additional simulation results}
\label{app:additional-sims}

This appendix complements the main simulation results in
Section 4.2 by providing a more detailed view of
time--unit--specific causal estimation performance across varying degrees of
latent factor mis-specification.

For each value of $\beta_U \in \{0, 1, 2, 5\}$, we consider the optimally
selected configuration $(K^\star,\widehat{\omega})$ and evaluate predictive
accuracy for $\tau_{\mathrm{ATT}}(t)$ exclusively on masked post-treatment
pseudo-treated observations (see Section~\ref{sec:panel_sim} for design 
details). We report squared bias $\big(\widehat{\tau}_{\mathrm{ATT}}(t) - \tau_{\mathrm{ATT}}(t)\big)^2$, and $95\%$ interval score,
both averaged across units that are masked at each time point $t$. These
figures provide a dynamic view of causal prediction quality beyond the global
averages shown in Figure 4 in the main text.

\begin{figure}[H]
    \centering
    \includegraphics[width=0.70\textwidth]{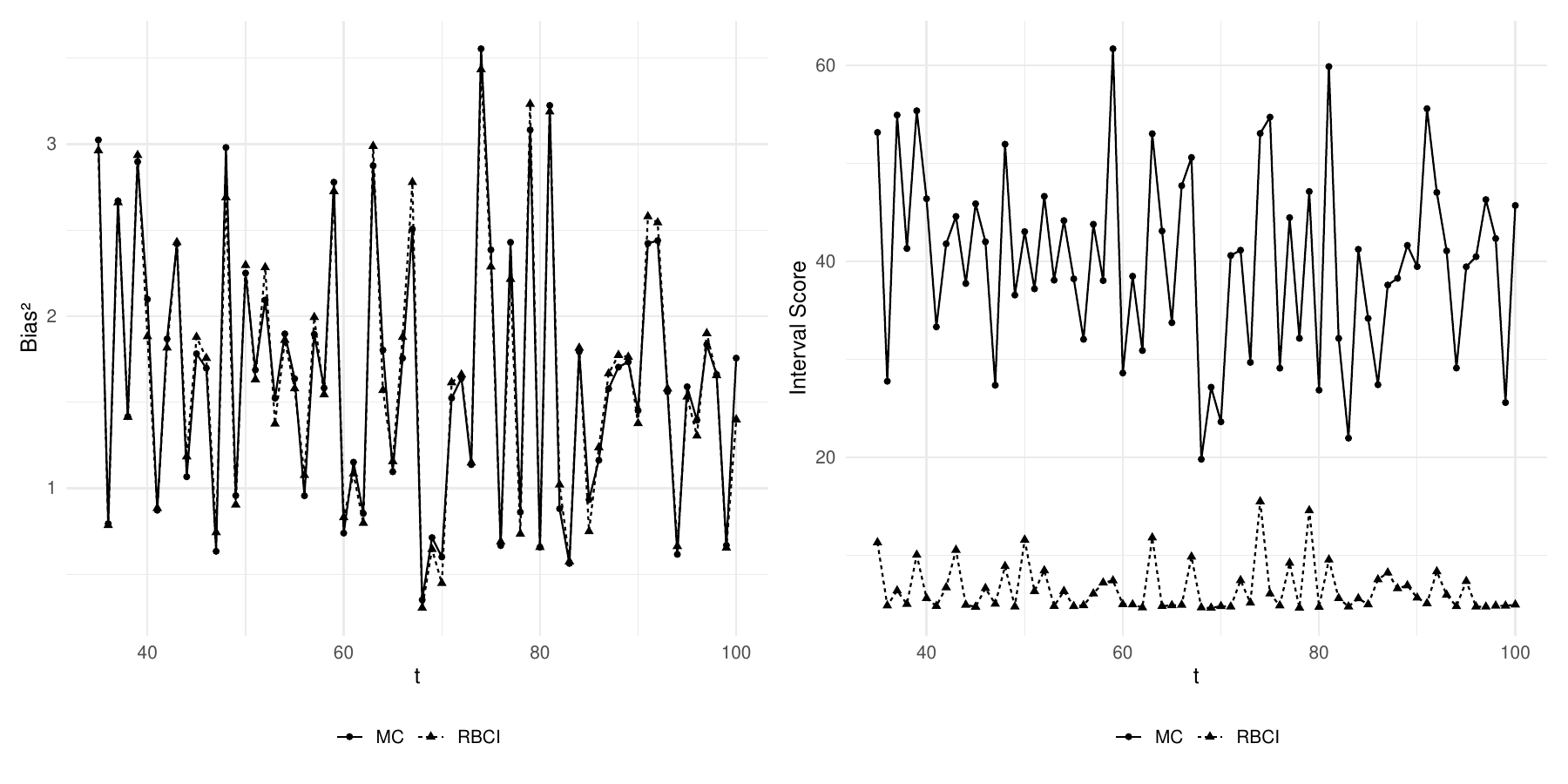}
    \caption{Time-varying squared bias (left) and interval score (right) under
    low latent factor mis-specification ($\beta_U=1$). Robust Bayesian causal inference (RBCI) yields consistently lower interval scores and improved point estimation relative to matrix completion (MC).}
    \label{fig:compare-low}
\end{figure}

\begin{figure}[H]
    \centering
    \includegraphics[width=0.70\textwidth]{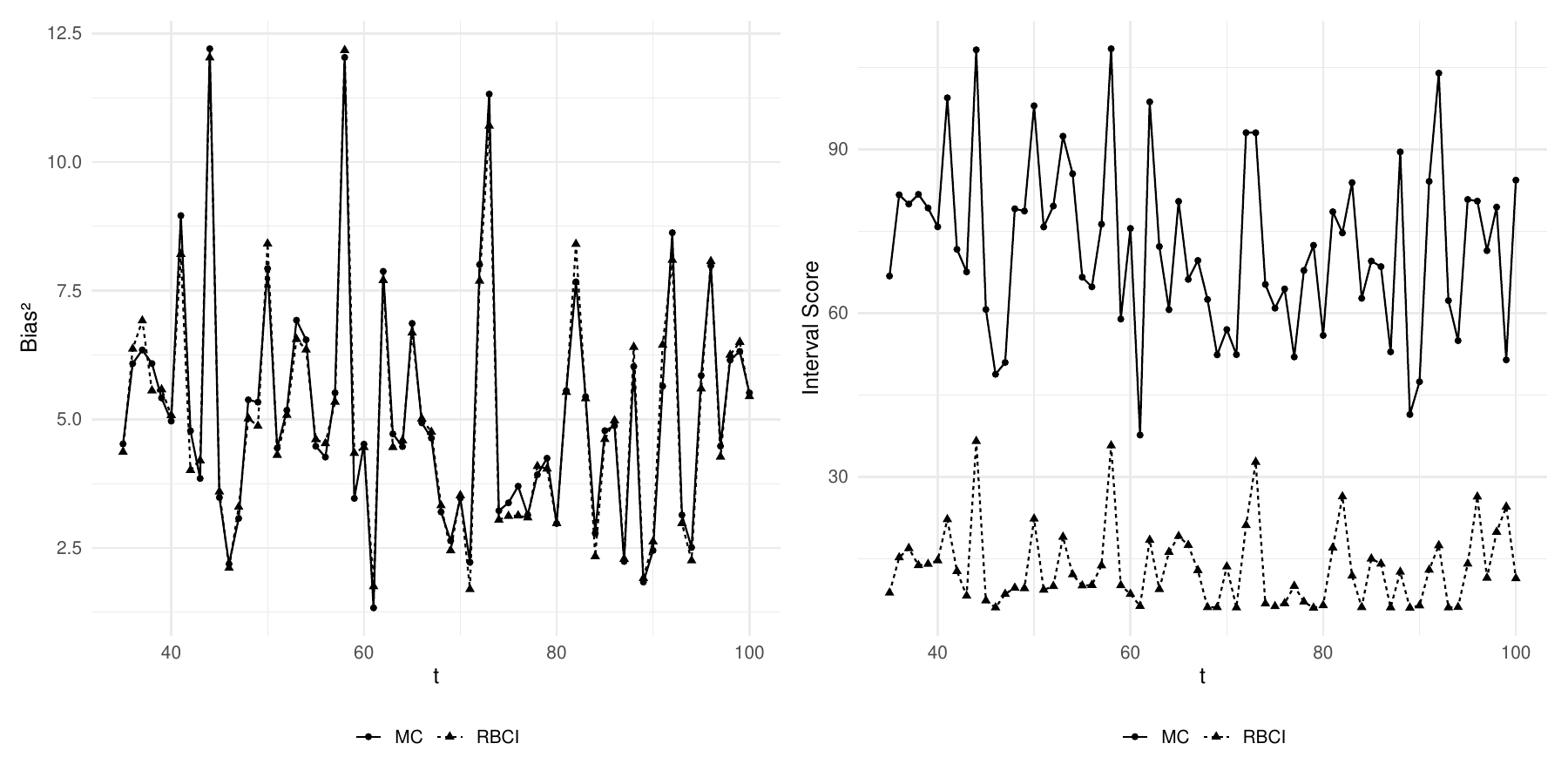}
    \caption{Medium latent factor mis-specification ($\beta_U=2$). Performance
    gains of robust Bayesian causal inference (RBCI) increase relative to the low mis-specification scenario against matrix completion (MC).}
    \label{fig:compare-medium}
\end{figure}

\begin{figure}[H]
    \centering
    \includegraphics[width=0.70\textwidth]{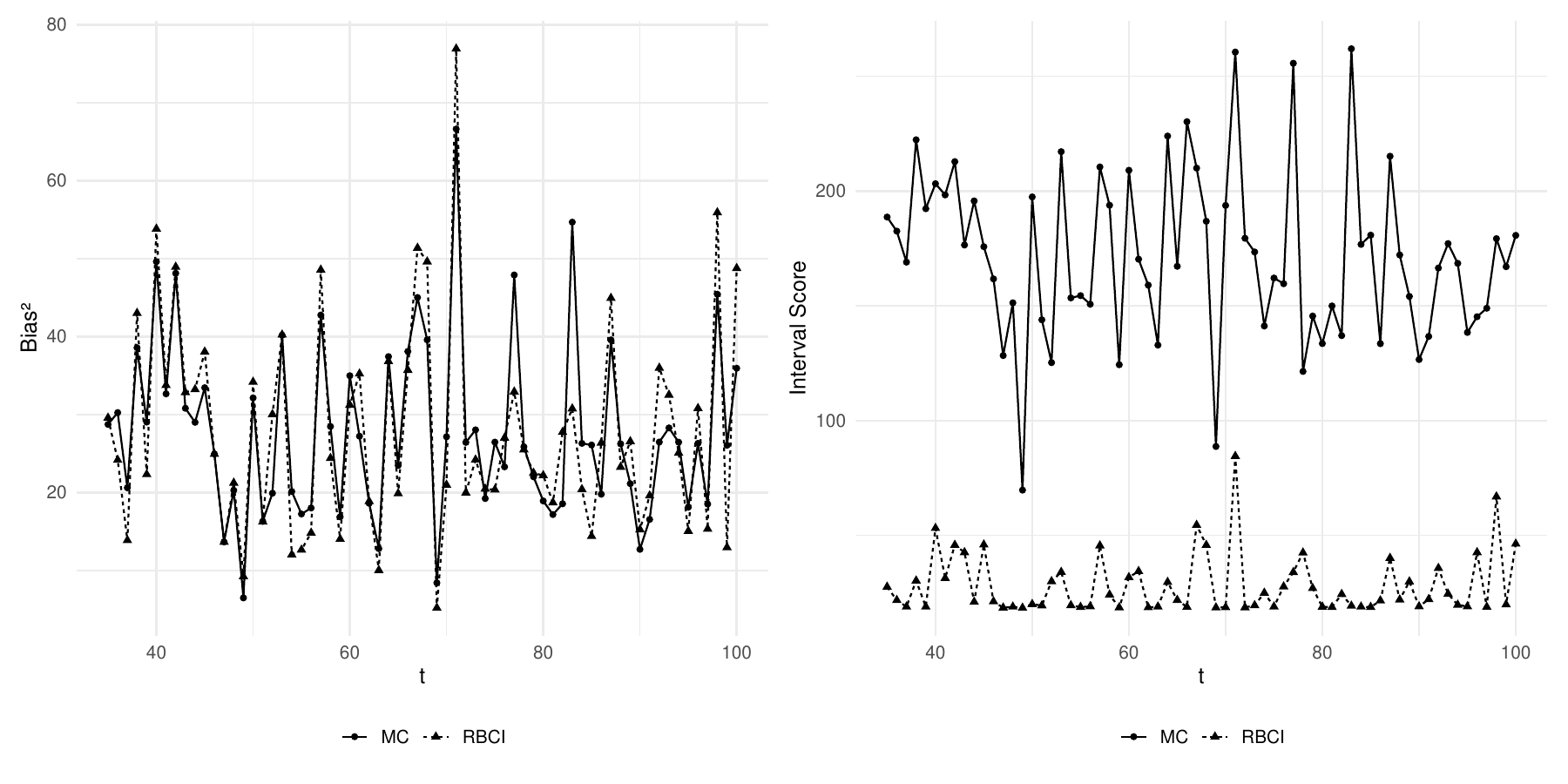}
    \caption{High latent factor mis-specification ($\beta_U=5$). Robust Bayesian causal inference (RBCI) shows reductions in squared bias and interval score compared with matrix completion (MC), highlighting robustness to severe unobserved confounding.}
    \label{fig:compare-high}
\end{figure}

Figures~\ref{fig:compare-low}--\ref{fig:compare-high} show that RBCI is able to
track the post-treatment dynamics with improved accuracy relative to matrix
completion (MC) when latent confounding is present. In the low and medium
mis-specification settings, RBCI achieves a uniformly lower interval score
and reduced squared bias across most post-treatment years. Under high
mis-specification these gains become substantial, reflecting the ability of
the selected learning rate $\widehat{\omega}<1$ to temper overconfident
likelihood contributions and improve predictive robustness. Importantly, RBCI does not degrade coverage or point accuracy at earlier
post-treatment times where latent confounding exerts weaker influence, further
supporting the benefits of adaptively tuning $(K,\omega)$.

\section{Real datasets details}

\subsection{California Smoking Data}

We provide further details on the California tobacco intervention dataset analysed in Section~\ref{subsec:Smoking}. The balanced panel includes annual per-capita cigarette consumption for 39 U.S.\ states from 1970 to 2000, with all outcomes log-transformed prior to analysis. California is treated as the intervention state following Proposition~99 in 1989, while the remaining 38 states serve as never-treated controls. To enable unbiased hyperparameter selection, we randomly mask 20\% of pre-intervention observations belonging to control states and evaluate predictive performance on these held-out entries. To assess causal prediction accuracy, we additionally construct a placebo evaluation set by masking post–1989 outcomes for a random 15\% subset of control states and treating those cells as pseudo-treated. Both masking steps strictly respect time ordering, ensuring valid causal evaluation. Figure~S1 displays the panel layout, showing that all states are observed in every year of the study period and highlighting the absence of missing observations outside the deliberately masked entries. Figure~S2 plots the average (log) cigarette sales in California and the corresponding mean across control states. The parallel pre-intervention trajectories and the clear divergence after 1989 emphasise the need for factor-based counterfactual estimation.

\begin{figure}[H]
  \centering
  \includegraphics[width=0.7\textwidth]{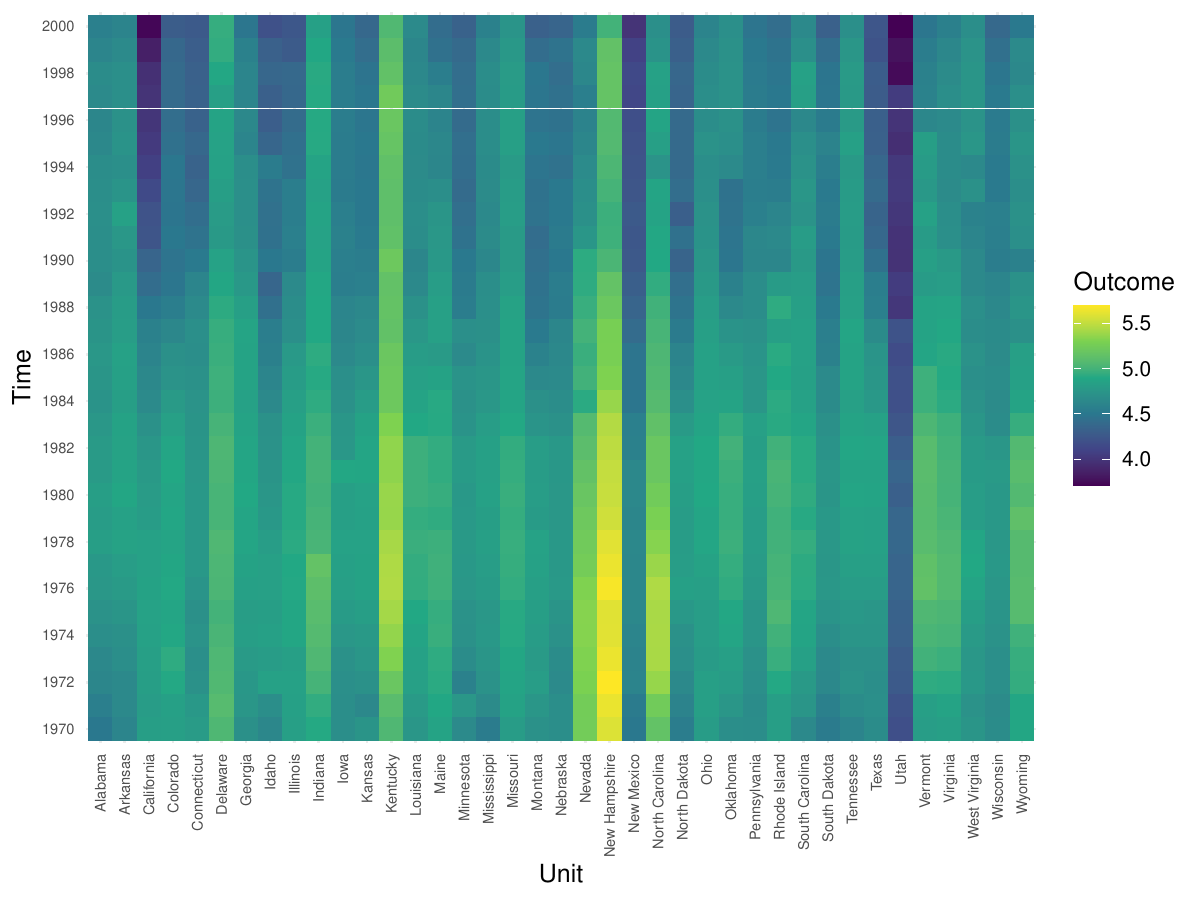}
  \caption{Panel structure for the California smoking dataset: log per-capita cigarette sales observed for 39 states over 1970--2000. California is the treated unit (red stripe at bottom); controls shown above.}
  \label{fig:prop99-heatmap}
\end{figure}

\begin{figure}[H]
  \centering
  \includegraphics[width=0.70\textwidth]{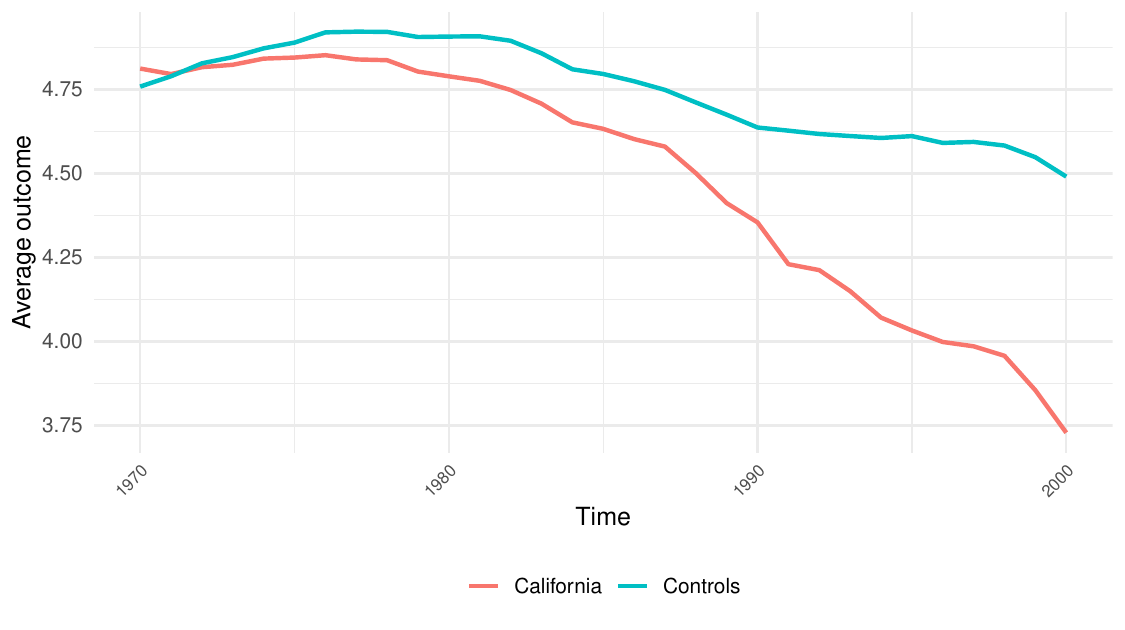}
  \caption{Average log per-capita cigarette sales for California and the mean across never-treated states, 1970--2000. Proposition~99 is enacted in 1989.}
  \label{fig:prop99-avg-ts}
\end{figure}

\subsection{French regional industrial policy dataset}

Here, we provide exploratory figures for the
French regional industrial policy dataset used in Section~\ref{subsec:France}.
The panel comprises annual observations for $N=148$ French employment areas
over $T=20$ years (1993--2012), with $13$ treated regions and $135$ never-treated
controls. We analyse three outcomes: firm entries, firm exits, and total
employment.

\begin{figure}[H]
\centering
\includegraphics[width=0.5\textwidth]{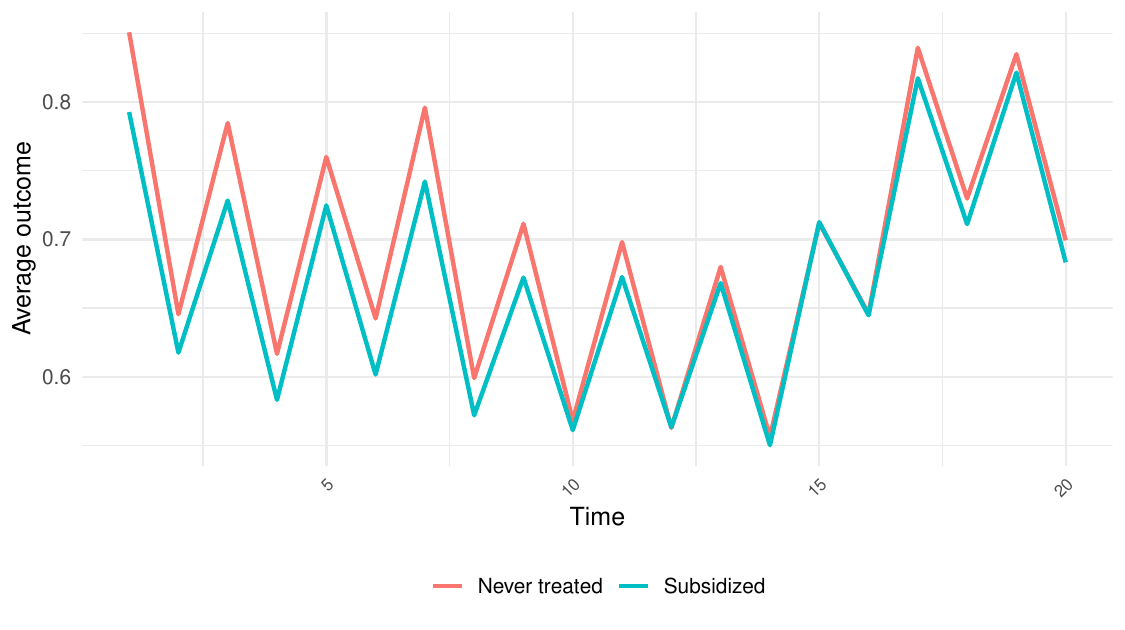}
\caption{Mean firm entry rates over time for subsidized against never-treated regions.}
\label{fig:fr-entries-ts}
\end{figure}

\begin{figure}[H]
\centering
\includegraphics[width=0.5\textwidth]{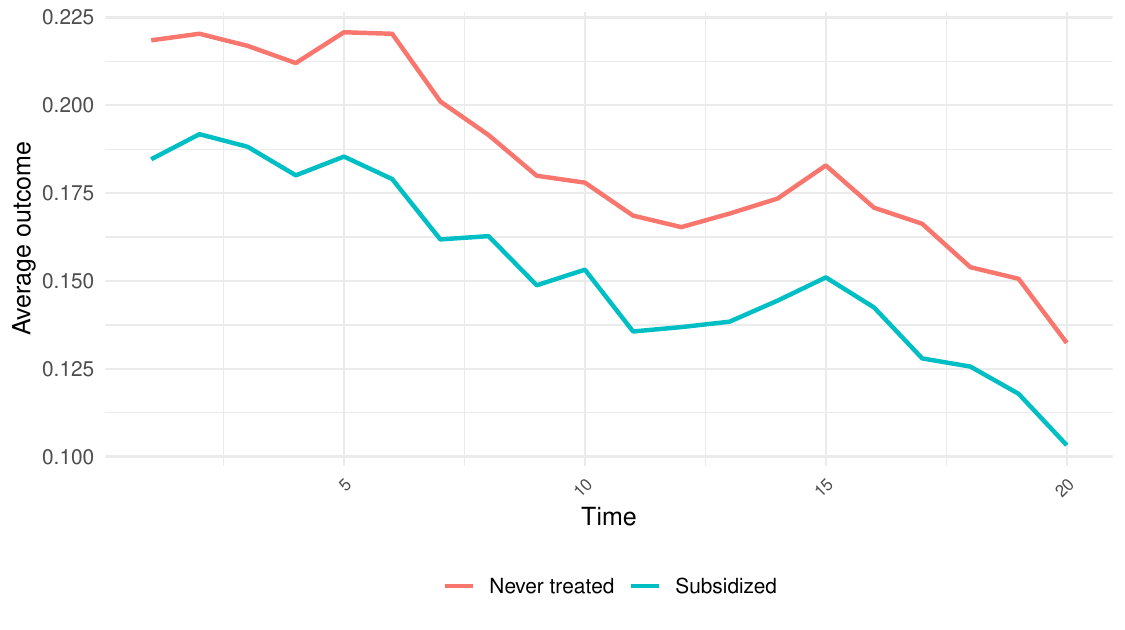}
\caption{Mean firm exit rates over time for subsidized vs.\ never-treated regions.}
\label{fig:fr-exits-ts}
\end{figure}

\begin{figure}[H]
\centering
\includegraphics[width=0.5\textwidth]{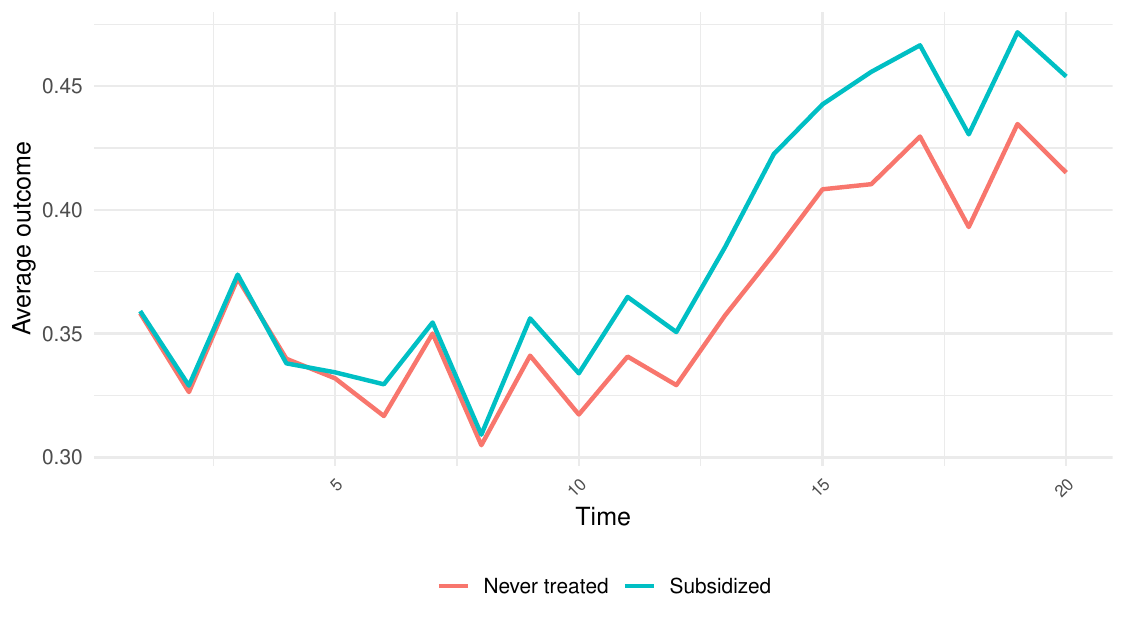}
\caption{Mean total employment over time for subsidized vs.\ never-treated regions.}
\label{fig:fr-employment-ts}
\end{figure}



\end{document}